\documentclass[12pt]{article}
\usepackage{graphics,graphicx}
\usepackage{amssymb,epsfig,amsmath,euscript,array}
\usepackage{cite}
\usepackage{axodraw4j}
\usepackage{pstricks}
\usepackage{color}

\makeatletter
\@addtoreset{equation}{section}
\makeatother

\newcounter{multieqs}

\newcommand{\be}{\begin{equation}}
\newcommand{\ee}{\end{equation}}

\newcommand{\bm}[1]{\mbox{\boldmath $#1$}}

\newcommand{\kslash}{k \!\!\! / }

\newcommand{\lslash}{l \!\! / }
\newcommand{\Pslash}{P \!\!\!\! / }

\newcommand{\islash}{i \!\!\! / }
\newcommand{\jslash}{j \!\!\! / }
\newcommand{\aslash}{a \!\!\! / }
\newcommand{\bslash}{{b \hspace{-6pt} \slash} }

\newcommand{\onslash}{1 \!\!\! / }
\newcommand{\twslash}{2 \!\!\!/ }
\newcommand{\thslash}{3 \!\!\!/ }
\newcommand{\foslash}{4 \!\!\! / }
\newcommand{\fislash}{5 \!\!\! / }

\newcommand{\mslash}{m \!\!\! / }

\def\bd{\begin{document}}
\def\ed{\end{document}}
\def\nn{\nonumber}
\def\bea{\begin{eqnarray}}
\def\eea{\end{eqnarray}}

\def\ab{(ijab)}
\def\ba{(ijba)}
\def\ijab{{\tr}_{+}(\islash\, \jslash\, \aslash \, \bslash)}
\def\ijba{{\tr}_{+}(\islash\, \jslash\, \bslash \, \aslash)}
\def\ijaP{{\tr}_{+}(\islash\, \jslash\, \aslash \, \Pslash)}
\def\ijPLa{{\tr}_{+}(\islash\, \jslash\, \Pslash_L \, \aslash)}
\def\ijaPL{{\tr}_{+}(\islash\, \jslash\, \aslash \, \Pslash_L)}
\def\ijPLza{{\tr}_{+}(\islash\, \jslash\, \Pslash_{L;z} \, \aslash)}
\def\ijaPLz{{\tr}_{+}(\islash\, \jslash\, \aslash \, \Pslash_{L;z})}
\def\ijPa{{\tr}_{+}(\islash\, \jslash\, \Pslash \, \aslash)}
\def\iaPb{{\tr}_{+}(\islash\, \aslash\, \Pslash \, \bslash)}
\def\ibPa{{\tr}_{+}(\islash\, \bslash\, \Pslash \, \aslash)}
\def\ijPmu{{\tr}_{+}(\islash\, \jslash\, \Pslash \, \mu)}
\def\ibmuP{{\tr}_{+}(\islash\, \bslash\, \mu \, \Pslash)}
\def\ibmua{{\tr}_{+}(\islash\, \bslash\, \mu \, \aslash)}
\def\iamub{{\tr}_{+}(\islash\, \aslash\, \mu \, \bslash)}
\def\jaPb{{\tr}_{+}(\jslash\, \aslash\, \Pslash \, \bslash)}
\def\ijmuP{{\tr}_{+}(\islash\, \jslash\, \mu \, \Pslash)}
\def\ijmum{{\tr}_{+}(\islash\, \jslash\, \mu \, \mslash)}
\def\ijmmu{{\tr}_{+}(\islash\, \jslash\, \mslash \, \mu)}
\def\ijmP{{\tr}_{+}(\islash\, \jslash\, \mslash \, \Pslash)}
\def\iabP{{\tr}_{+}(\islash\, \aslash\, \bslash \, \Pslash)}
\def\ijbP{{\tr}_{+}(\islash\, \jslash\, \bslash \, \Pslash)}
\def\jbPa{{\tr}_{+}(\jslash\, \bslash\, \Pslash \, \aslash)}
\def\ijPb{{\tr}_{+}(\islash\, \jslash\, \Pslash \, \bslash)}
\def\jbmua{{\tr}_{+}(\jslash\, \bslash\, \mu \, \aslash)}

\def\loablt{ {\tr}_{+}(\lslash_1\, \aslash \, \bslash\, \lslash_2)}

\def\ijlolt{{\tr}_{+}(\islash\, \jslash\, \lslash_1 \, \lslash_2)}
\def\ijltlo{{\tr}_{+}(\islash\, \jslash\, \lslash_2 \, \lslash_1)}
\def\ibloa{{\tr}_{+}(\islash\, \bslash\, \lslash_1 \, \aslash)}
\def\jaltb{{\tr}_{+}(\jslash\, \aslash\, \lslash_2 \, \bslash)}
\def\ialtb{{\tr}_{+}(\islash\, \aslash\, \lslash_2 \, \bslash)}
\def\bltloa{{\tr}_{+}(\bslash\, \lslash_2\, \lslash_1 \, \aslash)}
\def\jbloa{{\tr}_{+}(\jslash\, \bslash\, \lslash_1 \, \aslash)}
\def\ibPb{{\tr}_{+}(\islash\, \bslash\, \Pslash \, \bslash)}
\def\ijltb{{\tr}_{+}(\islash\, \jslash\, \lslash_2 \, \bslash)}

\def\ijloa{{\tr}_{+}(\islash\, \jslash\,  \lslash_1 \, \aslash)}
\def\ijblt{{\tr}_{+}(\islash\, \jslash\,  \bslash \, \lslash_2)}

\def\jakb{{\tr}_{+}(\jslash\, \aslash\, \kslash \, \bslash)}
\def\iakb{{\tr}_{+}(\islash\, \aslash\, \kslash \, \bslash)}

\def\tofo{{\tr}_{+}(\onslash\, \thslash\, \twslash \, \foslash)}
\def\foto{{\tr}_{+}(\onslash\, \thslash\, \foslash \, \twslash)}
\def\tofi{{\tr}_{+}(\onslash\, \thslash\, \twslash \, \fislash)}
\def\fito{{\tr}_{+}(\onslash\, \thslash\, \fislash \, \twslash)}

\def\lrangle#1#2{\langle #1\,#2\rangle}

\newcommand\mA{\mathcal{A}}
\newcommand\mB{\mathcal{B}}
\def\Li{{$\rm Li}_2$}
\def\eps{\epsilon}
\def\epsuv{{\epsilon_{\rm \mbox{\tiny UV}}}}
\let\bm=\bibitem
\let\la=\label

\def\npb#1#2#3{Nucl. Phys. {\bf{B#1}} #3 (#2)}
\def\plb#1#2#3{Phys. Lett. {\bf{#1B}} #3 (#2)}
\def\prl#1#2#3{Phys. Rev. Lett. {\bf{#1}} #3 (#2)}
\def\prd#1#2#3{Phys. Rev. {D \bf{#1}} #3 (#2)}
\def\cmp#1#2#3{Comm. Math. Phys. {\bf{#1}} #3 (#2)}
\def\cqg#1#2#3{Class. Quantum Grav. {\bf{#1}} #3 (#2)}
\def\nppsa#1#2#3{Nucl. Phys. B (Proc. Suppl.) {\bf{#1A}}#3 (#2)}
\def\ap#1#2#3{Ann. of Phys. {\bf{#1}} #3 (#2)}
\def\ijmp#1#2#3{Int. J. Mod. Phys. {\bf{A#1}} #3 (#2)}
\def\rmp#1#2#3{Rev. Mod. Phys. {\bf{#1}} #3 (#2)}
\def\mpla#1#2#3{Mod. Phys. Lett. {\bf A#1} #3 (#2)}
\def\jhep#1#2#3{J. High Energy Phys. {\bf #1} #3 (#2)}
\def\atmp#1#2#3{Adv. Theor. Math. Phys. {\bf #1} #3 (#2)}
\newcommand{\EQ}[1]{\begin{equation} #1 \end{equation}}
\newcommand{\AL}[1]{\begin{subequations}\begin{align} #1 \end{align}\end{subequations}}
\newcommand{\SP}[1]{\begin{equation}\begin{split} #1 \end{split}\end{equation}}
\newcommand{\ALAT}[2]{\begin{subequations}\begin{alignat}{#1} #2 \end{alignat}
                        \end{subequations}}
\def\beqa{\begin{eqnarray}}
\def\eeqa{\end{eqnarray}}
\def\beq{\begin{equation}}
\def\eeq{\end{equation}}
\def\sst{\scriptscriptstyle}
\def\thetabar{\bar\theta}
\def\Tr{{\rm Tr}}
\def\one{\mbox{1 \kern-.59em {\rm l}}}
 \def\Nh{\hat{N}}

\newcommand{\half}{{\textstyle {1 \over 2}}}

\def\a{\alpha}      \def\da{{\dot\alpha}}
\def\b{\beta}       \def\db{{\dot\beta}}
\def\c{\gamma}  \def\G{\Gamma}  \def\cdt{\dot\gamma}
\def\d{\delta}  \def\D{\Delta}  \def\ddt{\dot\delta}
\def\e{\epsilon}        \def\vare{\varepsilon}
\def\f{\phi}    \def\F{\Phi}    \def\vvf{\f}
\def\h{\eta}
\def\k{\kappa}
\def\l{\lambda} \def\L{\Lambda}
\def\m{\mu} \def\n{\nu}
\def\o{\omega}
\def\p{\pi} \def\P{\Pi}
\def\r{\rho}
\def\s{\sigma}  \def\S{\Sigma}
\def\t{\tau}
\def\th{\theta} \def\Th{\Theta} \def\vth{\vartheta}
\def\X{\Xeta}
\def\z{\zeta}
\def\de{\partial}

\def\cA{{\cal A}} \def\cB{{\cal B}} \def\cC{{\cal C}}
\def\cD{{\cal D}} \def\cE{{\cal E}} \def\cF{{\cal F}}
\def\cG{{\cal G}} \def\cH{{\cal H}} \def\cI{{\cal I}}
\def\cJ{{\cal J}} \def\cK{{\cal K}} \def\cL{{\cal L}}
\def\cM{{\cal M}} \def\cN{{\cal N}} \def\cO{{\cal O}}
\def\cP{{\cal P}} \def\cQ{{\cal Q}} \def\cR{{\cal R}}
\def\cS{{\cal S}} \def\cT{{\cal T}} \def\cU{{\cal U}}
\def\cV{{\cal V}} \def\cW{{\cal W}} \def\cX{{\cal X}}
\def\cY{{\cal Y}} \def\cZ{{\cal Z}}

\def\ua{\underline{\alpha}}
\def\ub{\underline{\phantom{\alpha}}\!\!\!\beta}
\def\uc{\underline{\phantom{\alpha}}\!\!\!\gamma}
\def\um{\underline{\mu}}
\def\ud{\underline\delta}
\def\ue{\underline\epsilon}
\def\una{\underline a}\def\unA{\underline A}
\def\unb{\underline b}\def\unB{\underline B}
\def\unc{\underline c}\def\unC{\underline C}
\def\und{\underline d}\def\unD{\underline D}
\def\une{\underline e}\def\unE{\underline E}
\def\unf{\underline{\phantom{e}}\!\!\!\! f}\def\unF{\underline F}
\def\unm{\underline m}\def\unM{\underline M}
\def\unn{\underline n}\def\unN{\underline N}
\def\unp{\underline{\phantom{a}}\!\!\! p}\def\unP{\underline P}
\def\unq{\underline{\phantom{a}}\!\!\! q}
\def\unQ{\underline{\phantom{A}}\!\!\!\! Q}
\def\unH{\underline{H}}

\def\As {{A \hspace{-6.4pt} \slash}\;}
\def\bs {{b \hspace{-6.4pt} \slash}\;}
\def\Ds {{D \hspace{-6.4pt} \slash}\;}
\def\ds {{\del \hspace{-6.4pt} \slash}\;}
\def\ss {{\s \hspace{-6.4pt} \slash}\;}
\def\ks {{ k \hspace{-6.4pt} \slash}\;}
\def\ps {{p \hspace{-6.4pt} \slash}\;}
\def\pas {{{p_1} \hspace{-6.4pt} \slash}\;}
\def\pbs {{{p_2} \hspace{-6.4pt} \slash}\;}
\def\Ps {{P \hspace{-6.4pt} \slash}\;}
\def\Qs {{Q \hspace{-6.4pt} \slash}\;}

\def\Fh{\hat{F}}
\def\Vh{\hat{V}}
\def\Xh{\hat{X}}
\def\ah{\hat{a}}
\def\xh{\hat{x}}
\def\yh{\hat{y}}
\def\ph{\hat{p}}
\def\xih{\hat{\xi}}
\def\psit{\tilde{\psi}}
\def\Psit{\tilde{\Psi}}
\def\tht{\tilde{\th}}
\def\lt{\tilde{\lambda}}
\def\hl{\hat{\lambda}}
\def\hlt{\hat{\tilde{\lambda}}}
\def\llt{\tilde{l}}
\def\At{\tilde{A}}
\def\Qt{\tilde{Q}}
\def\Rt{\tilde{R}}
\def\Nt{\tilde{N}}

\def\at{\tilde{a}}
\def\st{\tilde{s}}
\def\ft{\tilde{f}}
\def\pt{\tilde{p}}
\def\qt{\tilde{q}}
\def\vt{\tilde{v}}
\def\nt{\tilde{n}}

\def\delb{\bar{\partial}}
\def\bz{\bar{z}}
\def\bD{\bar{D}}
\def\bB{\bar{B}}

\def\bk{{\bf k}}
\def\bl{{\bf l}}
\def\bp{{\bf p}}
\def\bq{{\bf q}}
\def\br{{\bf r}}
\def\bx{{\bf x}}
\def\by{{\bf y}}
\def\bR{{\bf R}}
\def\bV{{\bf V}}

\def\d{\delta}\def\D{\Delta}\def\ddt{\dot\delta}
\def\pa{\partial} \def\del{\partial}
\def\xx{\times}
\def\uno{\mbox{1 \kern-.59em {\rm l}}}
\def\trp{^{\top}}
\def\inv{^{-1}}
\def\dag{{^{\dagger}}}
\def\pr{^{\prime}}
\def\lan{\langle}
\def\ran{\rangle}
\def\rar{\rightarrow}
\def\lar{\leftarrow}
\def\lrar{\leftrightarrow}
\newcommand{\0}{\,\!}      
\def\one{1\!\!1\,\,}
\def\im{\imath}
\def\jm{\jmath}
\newcommand{\tr}{\mbox{tr}}
\newcommand{\slsh}[1]{/ \!\!\!\! #1}
\def\vac{|0\rangle}
\def\lvac{\langle 0|}
\def\hlf{\frac{1}{2}}
\def\ove#1{\frac{1}{#1}}
\def\Box{\square}
\def\ZZ{\mathbb{Z}}
\def\CC#1{({\bf #1})}
\def\bcomment#1{}
\def\bfhat#1{{\bf \hat{#1}}}
\def\VEV#1{\left\langle #1\right\rangle}
\newcommand{\ex}[1]{{\rm e}^{#1}} \def\ii{{\rm i}}
\def\rr{{\rm r}} \def\rs{{\rm s}}\def\rv{{\rm v}}
\def\ri{{\rm i}}\def\rj{{\rm j}}
\newcommand{\lrbrk}[1]{\left(#1\right)}
\newcommand{\sfrac}[2]{{\textstyle\frac{#1}{#2}}}

\def\Li{{\rm Li}_2}

\font\mybb=msbm10 at 12pt
\def\bb#1{\hbox{\mybb#1}}

\font\myBB=msbm10 at 18pt
\def\BB#1{\hbox{\myBB#1}}

\begin{document}

\begin{flushright}
QMUL-PH-10-18
\end{flushright}

\vspace{20pt}

\begin{center}

{\Large \bf Tree-Level Formalism  }
\vspace{32pt}

{\mbox {\bf Andreas Brandhuber, Bill Spence and Gabriele Travaglini}}%
\footnote{
{\sffamily \{\tt a.brandhuber, w.j.spence, g.travaglini\}@qmul.ac.uk }}

{\em Centre for Research in String Theory\\
School of Physics\\
Queen Mary University of London\\
Mile End Road, London, E1 4NS\\
United Kingdom
 }

\vspace{30pt} {\bf Abstract}

\end{center}

\noindent
We review two novel techniques used to calculate tree-level scattering amplitudes efficiently: MHV diagrams, and on-shell recursion relations. For the MHV diagrams, we consider applications to tree-level amplitudes and focus in particular on the $\cN=4$ supersymmetric formulation. We also briefly describe the derivation of loop amplitudes using MHV diagrams. For the recursion relations, after presenting their general proof, we discuss several applications to massless theories with and without supersymmetry, to theories with massive particles, and to graviton amplitudes in General Relativity.  
This article is an invited review for a special issue of Journal of Physics A devoted to ÒScattering Amplitudes in Gauge TheoriesÓ.

\setcounter{page}{0}
\thispagestyle{empty}
\newpage


\section{Introduction   }
\setcounter{footnote}{0}

There has been rapid progress since 2004 in the study of new structures in scattering amplitudes in diverse quantum field theories, and their relationships with string theory and twistor theory. This has frequently come
about via sudden and surprising insights into new properties, or new understandings of known properties, of tree amplitudes -- a subject that one might have been expected to have been fully understood.

The initial work, following Witten's pioneering investigations into the twistor space localisation of amplitudes \cite{witten}, studied further the localisation of tree amplitudes. An early and vital insight was
that one could glue together MHV tree amplitudes in order to obtain non-MHV amplitudes \cite{csw1}; this expansion of tree amplitudes in terms of MHV diagrams provided a very natural way of understanding localisation on collections of lines in twistor space for more general tree amplitudes.%
\footnote{Note that in the context of twistor string theory the MHV diagrams are naturally related to the so-called disconnected prescription, while Witten's original approach \cite{witten}, further studied and extended in subsequent papers \cite{rsv1,rv,rsv2}, was termed the connected prescription. Both prescriptions have been shown to be equivalent in \cite{gmn}, where also intermediate prescriptions were identified, see also \cite{Bena:2004ry}.}
This  \lq\lq MHV diagram\rq\rq\ approach was then applied successfully to one-loop amplitudes in $\cN=4$ \cite{bst1}, $\cN=2$, $\cN=1$ \cite{quig, bbst1} and
even pure Yang-Mills theories \cite{bbst2}. 
Important steps towards the understanding of the MHV diagram approach arose from applying a particular non-local field redefinition to the lightcone Yang-Mills action \cite{Mansfield:2005yd, Gorsky:2005sf}. Extensions to non-MHV loop amplitudes have proved more difficult and this has impeded progress in this direction.
More recent work has found very interesting and related new structures in the integrands of loop amplitudes, by reformulating them in momentum twistor space or dual momentum space. Some of these topics are described elsewhere in this volume.

A second, and particularly striking new discovery was that of recursion relations for gauge theory at tree level \cite{bcfrec,bcfw}. These followed from a beautiful and  elegant analysis of the complex analytic structure of tree amplitudes, together with the basic physical principle of unitarity of the S-matrix \cite{S}. It was surprising that these recursion relations had not been found previously;  they epitomise a concrete realisation of the old S-matrix programme, by providing a method to compute S-matrix elements directly from the analytic properties of the amplitudes, without ever referring  to a Lagrangian, and requiring only on-shell quantities as input. These recursion relations have proved to be a very powerful and efficient tool as one may build more complex amplitudes from simpler ones. They have also been successfully extended to cover a wide class of field theories, including gravity \cite{bbstrec,cs} and also loop amplitudes \cite{Bern:2005hs,Bern:2005ji,Bern:2005cq}, enabling new results to be derived and new structures to be found.
Since they follow from basic properties of quantum field theory, interesting new applications continue to be found -- 
recent examples include recursion relations for loop integrands in $\cN=4$ super Yang-Mills (SYM) \cite{abcct, Boels:2010nw} and recursion relations for partially off-shell quantities such as form factors \cite{Brandhuber:2010ad} -- and it is to be expected that these techniques will hereon be an essential part of the armoury available for tackling problems in quantum field theories.

The present chapter of this review is organised as follows:
In the next  section we will review the approach to tree-level  amplitudes based on   MHV diagrams, describing 
the method  with some simple examples, as well as describing how it can be applied  at one loop. In Section 3 we turn to recursion relations, with examples of how they can be used in gauge theories with massless and massive particles and in gravity. Finally we describe the manifestly supersymmetric formulation in $\cN=4$ SYM using Nair's on-shell superspace \cite{Nair}.

\setcounter{footnote}{0}

\section{The MHV diagram method}
\label{MHVsection}

The key quantity in the MHV diagram method is the  MHV scattering amplitude of $n$ gluons,   given by 
\cite{Parke:1986gb,Berends:1987cv,Mangano:1987xk,Berends:1987me}
\beq
\label{MHV}
A_{\rm MHV } (1^+, 2^+, \ldots , i^- , \ldots , j^- , \ldots, n^+) =  
i(2 \pi)^4 g^{n-2} \delta^{(4)} \big(\sum_{i=1}^n \l_i \lt_i \big) \, 
{\lan ij\ran^4 \over \lan 12\ran \lan23\ran \cdots \lan n1\ran}
\ . 
\eeq
A remarkable feature of this strikingly compact expression is that it is a holomorphic function of the spinor variables of the $n$ scattered gluons. In \cite{witten}, Witten exploited this fact to perform a Fourier transform of this amplitude to Penrose's twistor space, 
\beq
\label{local}
\int\!\!\prod_{i=1}^n d^2 \lt_i \, e^{i [\mu_i \lt_i] } A_{\rm MHV }   \ = \ 
A_{\rm PT} \int\!\!d^4x \, \prod_{i=1}^n \delta^{(2)} (\mu^{\dot{a}}_i+ x^{\dot{a} a } \l_{i, a})  
\ , 
\eeq
where 
\beq
A_{\rm PT} := {\lan ij\ran^4 \over \lan 12\ran \lan23\ran \cdots \lan n1\ran}
\ . 
\eeq
The right-hand side of  (\ref{local}) vanishes unless each  twistor coordinate of the $n$ gluons $\l_{a, i}, \mu_{\dot{a}, i}$
satisfies the relation 
\beq
\mu^{\dot{a}} + x^{\dot{a} a} \lambda_a \, = \, 0, \qquad \dot{a}=1,2
\ . 
\eeq
This is the equation of a line in twistor space; in twistor theory, this is a famous relation -- the incidence relation, which associates points in conformally compactified, complexified Minkowski space to complex lines, or 
$\bb{C}\bb{P}^1$'s in twistor space  \cite{penrose}. 
Keeping this observation in mind, Cachazo, Svr\v{c}ek and Witten (CSW) suggested in \cite{csw1} that one may think of an MHV amplitude as an effective, {\it local} interaction in spacetime.%
\footnote{Mansfield showed in \cite{Mansfield:2005yd} that the MHV rules, to be explained below, can be derived at tree level by a canonical change of variables in the Yang-Mills path integral. The resulting action is local (only) in lightcone time. We briefly review this approach in Section \ref{cov}.}  
The idea is to introduce a particular off-shell continuation of an MHV amplitude, the MHV vertex. By gluing these MHV vertices in an appropriate way, to be discussed momentarily, we will construct amplitudes with an arbitrary number $q$ of negative helicities -- what we will call an ${\rm N}^{q-2}{\rm MHV}$ amplitude. 

We can easily count the number of vertices $v$ that are needed in order to calculate ${\rm N}^{q-2}{\rm MHV}$ amplitude. 
$v$ MHV vertices provide us with $2v$ negative helicities, since each vertex has the MHV helicity configuration.  
At $l$ loops, a generic diagram has $v-1+l$ internal lines, each of which absorbs exactly one negative helicity. 
This is because a propagator connects an outgoing gluon of a certain helicity on one MHV vertex with an outgoing gluon of the opposite helicity on another MHV vertex.
We are thus left with $q= v+1-l$ negative helicities, and hence 
\beq 
v=q-1+l
\ . 
\eeq
At tree level ($l=0$) we will then draw all possible MHV diagrams with $q-1$ MHV vertices and ``dress" them with the external gluons in such a way that each vertex has the MHV helicity configuration. 

Now two important issues have to be addressed in order to build up the diagrams. Firstly, vertices must be connected with an appropriate propagator. Secondly, \eqref{MHV}  gives the expression for an {\it on-shell amplitude}, and, therefore, one has to provide an off-shell continuation of such an amplitude in order to use it as a {\it vertex}. In   \cite{csw1}, CSW provided us with answers to both questions, as we now describe. 

\subsection{Propagators and off-shell continuation}
\label{section-offshell}
The answer to the first question just posed is extremely simple -- the internal propagators connecting MHV vertices are simply scalar propagators, $i/ (L^2 + i \varepsilon)$. This result can be justified from the Lagrangian formulation of MHV rules described in \cite{Mansfield:2005yd}.
Next, we consider the issue of off-shell continuation. 
Let $L$ be an internal (hence off-shell) momentum. 
The MHV amplitude depends on the holomorphic spinors associated to the external, on-shell particles, and we will now provide a prescription that associates a spinor to the internal leg with (off-shell)  momentum $L$.  To this end,  
one introduces  an arbitrary lightlike reference vector $\eta_{\a \da} = \eta_{\a} \tilde{\eta}_{\da}$. Then,  there is a unique way to decompose $L$ as 
 \cite{Bena:2004ry,Kosower:2004yz}
\beq
\label{offshell}
L \  = \ l \, + \, z \eta
\ , 
\eeq
where $l:= \l_L \lt_L$ is a null momentum and $z$ is a real parameter, which can easily be seen to be equal to $z=L^2/(2 L\cdot \eta)$. The off-shell continuation proposed in \cite{csw1} consists in using the spinor $\l_L$ as the off-shell continuation for the internal leg of momentum $L$. It then follows that 
\beqa
\label{off1}
\l_{L, \a} & = & {L_{\a \da} \tilde{\eta}^{\da}
\over [ \lt_L \, \tilde{\eta}]
}
\ .
 \eeqa
These equations coincide with the  prescription originally proposed in \cite{csw1} by CSW for
determining the spinor variable   $\l_L$ associated with
the off-shell (i.e.~non-null) four-vector $L$ defined in
\eqref{offshell}. The denominators arising from the right-hand side of \eqref{off1}
will be irrelevant for our applications, since each MHV diagram is invariant under rescalings of the internal spinor variables; hence we will discard them and simply replace $\l_{L, \a} \to  L_{\a \da} \tilde{\eta}^{\da}$. 
Each MHV diagram will therefore depend on the particular choice of $\tilde{\eta}$; one will then have to prove that summing over all MHV diagrams, this spurious dependence cancels, showing that the MHV diagrams produce covariant answers. In \cite{csw1} this was originally shown for diagrams involving one propagator i.e. NMHV gluon amplitudes.
Later an elegant proof was found in \cite{Risager:2005vk} where it was shown that MHV diagrams are in direct correspondence with on-shell recursion relations (which by construction lead to covariant answers and will be described in later sections) where all negative helicity gluons are shifted. This was later extended to include all tree amplitudes in $\cN=4$ SYM \cite{Elvang:2008vz}.

\subsection{Examples of application} 
\label{rara}

The proposal of  \cite{csw1} that tree amplitudes could be built by gluing together MHV tree amplitudes as vertices, joining them with propagators, was very intuitive and gave an answer to the basic question of how tree amplitudes for $n$-particle scattering could be relatively simple,
when they were obtained by summing a rapidly increasing number of Feynman diagrams.
This method promised a new, efficient means for calculating amplitudes, seemingly sidestepping the complications of gauge fixing and dealing with unwieldy sets of diagrams.

We will start with some simple applications at tree level and defer
the more complicated example of all NMHV tree amplitudes for later
where we will combine the CSW rules with Nair's on-shell superspace
formalism.

As a warm-up we will now rederive the well-known fact that the four-point
tree amplitude $\langle 1^+ 2^- 3^- 4^-\rangle$ vanishes. More generally, all trees with equal helicity gluons or only a single gluon of opposite helicity are zero. A very important exception to this are the three-point amplitudes $\langle 1^+ 2^- 3^- \rangle$ and $\langle 1^- 2^+ 3^+ \rangle$ which vanish for real momenta in Minkowski signature but are non-vanishing for complex momenta (as we will see later, this has profound consequences). Formally, our example is a so-called next-to-MHV (NMHV) amplitude, as it has three negative helicity gluons.

\begin{figure}[h]
\begin{center}
\scalebox{0.55}
{
\fcolorbox{white}{white}{
  \begin{picture}(468,230) (63,-43)
    \SetWidth{1.0}
    \SetColor{Black}
    \Line[arrow,arrowpos=0.5,arrowlength=5,arrowwidth=2,arrowinset=0.2](128,70)(64,134)
    \Line[arrow,arrowpos=0.5,arrowlength=5,arrowwidth=2,arrowinset=0.2](128,70)(64,6)
    \Text(176,54)[lb]{\Large{\Black{$P$}}}
    \Text(144,86)[lb]{\Large{\Black{$-$}}}
    \Text(208,86)[lb]{\Large{\Black{$+$}}}
    \Text(64,150)[lb]{\Large{\Black{$p_2^-$}}}
    \Text(288,150)[lb]{\Large{\Black{$p_3^-$}}}
    \Text(288,-10)[lb]{\Large{\Black{$p_4^-$}}}
    \Text(64,-10)[lb]{\Large{\Black{$p_1^+$}}}
    \Line[arrow,arrowpos=0.5,arrowlength=5,arrowwidth=2,arrowinset=0.2](416,118)(352,182)
    \Line[arrow,arrowpos=0.5,arrowlength=5,arrowwidth=2,arrowinset=0.2](416,22)(352,-42)
    \Line[arrow,arrowpos=0.5,arrowlength=5,arrowwidth=2,arrowinset=0.2](416,118)(480,182)
    \Line[arrow,arrowpos=0.5,arrowlength=5,arrowwidth=2,arrowinset=0.2](416,22)(480,-42)
    \Line[arrow,arrowpos=0.5,arrowlength=5,arrowwidth=2,arrowinset=0.2](128,70)(224,70)
    \Line[arrow,arrowpos=0.5,arrowlength=5,arrowwidth=2,arrowinset=0.2](224,70)(288,134)
    \Line[arrow,arrowpos=0.5,arrowlength=5,arrowwidth=2,arrowinset=0.2](224,70)(288,6)
    \Line[arrow,arrowpos=0.5,arrowlength=5,arrowwidth=2,arrowinset=0.2](416,22)(416,118)
    \Text(432,70)[lb]{\Large{\Black{$P'$}}}
    \Text(400,38)[lb]{\Large{\Black{$-$}}}
    \Text(400,102)[lb]{\Large{\Black{$+$}}}
    \Text(336,-26)[lb]{\Large{\Black{$p_1^+$}}}
    \Text(336,166)[lb]{\Large{\Black{$p_2^-$}}}
    \Text(496,166)[lb]{\Large{\Black{$p_3^-$}}}
    \Text(496,-26)[lb]{\Large{\Black{$p_4^-$}}}
  \end{picture}
}}
\end{center}
\caption{\it The two MHV diagrams contributing to $\langle 1^+ 2^- 3^- 4^-\rangle$}
\label{4ptnmhvdiag}
\end{figure}
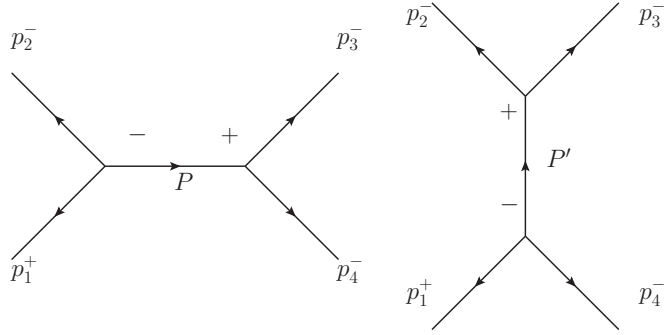

There are two diagrams contributing to this amplitude, where two three-point MHV vertices are connected by a propagator, see Figure \ref{4ptnmhvdiag}.
Using the fact that $P^2 = (p_1+p_2)^2 = \langle 1 2 \rangle [21]$
and $P'^2 = (p_1+p_4)^2 = \langle 1 4 \rangle [41]$ we obtain for the
sum of the two diagrams, 
\beq\label{4ptnmhw}
i^3 \frac{\langle 2 P\rangle^3}{\langle P 1\rangle\langle 1 2\rangle} \frac{1}{\langle 1 2 \rangle [21]} \frac{\langle 34\rangle^3}{\langle4 \, -\!\!P\rangle \langle-\!P 3\rangle} +
i^3 \frac{\langle 2 3\rangle^3}{\langle 3 -\!\!P' \rangle \langle -\!P' 2\rangle} \frac{1}{\langle 1 4 \rangle [41]} \frac{\langle P'4\rangle^3}{\langle 4 1\rangle \langle 1 P'\rangle} \, .
\eeq
Now we have to use the off-shell continuation introduced earlier namely
$\lambda_{P,\alpha} = P_{\alpha\dot{\alpha}} \tilde{\eta}^{\dot{\alpha}}=
- i \lambda_{-\!P,\alpha}$, 
and we have denoted incoming internal momenta with an additional
minus sign since they correspond to an outgoing momentum $-P$. In our convention for on-shell  momenta,  
$p_{\alpha\dot\alpha} := \lambda_\alpha \tilde{\lambda}_{\dot\alpha}$,  whenever we
flip the orientation from outgoing to incoming we multiply $\lambda$ and $\tilde{\lambda}$ with $i$. Taking into account these extra factors of $i$ and using the off-shell continuation we get, for the first diagram%
\footnote{Note that, as we mentioned earlier,  we are allowed to drop the normalisation constant in the definition of the off-shell continuation since \eqref{mhv4ptdiag1} is homogeneous in the internal off-shell spinors -- this is a generic feature of MHV diagrams.}
\beq\label{mhv4ptdiag1}
i\frac{\langle 2 | P | \eta]^3 \langle 3 4 \rangle^3}{[\eta|P|1\rangle \langle 1 2 \rangle^2 [21] \langle 4 | P | \eta ] [ \eta | P |3 \rangle}
= i \frac{\langle 3 4 \rangle}{[21]} \frac{[1 \eta]^3}{[2 \eta][3 \eta][4 \eta]} \, ,
\eeq
where we used  $P = -p_1-p_2=p_3+p_4$ and identities such as
$\langle 2 | P | \eta] = \langle 2 | -p_1-p_2 | \eta]=
-\langle 2 1 \rangle [1 \eta] - \underbrace{\langle 2 2 \rangle}_{=0} [2 \eta] = \langle 1 2 \rangle [1 \eta]$. 
The evaluation of the second diagram gives
\beq\label{mhv4ptdiag2}
i \frac{\langle 1 4 \rangle}{[23]} \frac{[1 \eta]^3}{[2 \eta][3 \eta][4 \eta]} \, .
\eeq
The sum of \eqref{mhv4ptdiag1} and \eqref{mhv4ptdiag2} vanishes, since
\beq\label{4ptmhvzero}
\frac{\langle34\rangle}{[21]}+\frac{\langle14\rangle}{[23]}=\frac{[23]\langle34\rangle+[21]\langle14\rangle}{[21][23]}=
\frac{[2|p_3+p_1|4\rangle}{[21][23]}=
-\frac{[2|p_2+p_4|4\rangle}{[21][23]}=0 \ ,   
\eeq
confirming that the $\langle 1^+ 2^- 3^- 4^-\rangle$ amplitude is indeed zero.

This procedure can be easily generalised to calculate arbitrary $n$-point
amplitudes with $k$ negative helicity gluons. Such N$^{k-2}$MHV amplitudes
require one to sum all possible MHV diagrams with $k-1$ MHV vertices connected
by scalar propagators. Applications are not restricted to amplitudes with
gluons only but can also involve fermions or scalars \cite{GK, GGK}.

An obvious next step was to see if this method could be used to calculate loop amplitudes and we would like to briefly describe how this is achieved \cite{bst1}.
Much was already known about loop amplitudes, particularly after work using unitarity methods in 
the 1990's 
\cite{bddk, fusing,bm,dixon,Bern:1996je,bdk96}. Applications of modern unitarity techniques are discussed in great detail in the chapters of this review by Britto 
\cite{britto} and Bern and Huang \cite{zvi}.

The first and simplest example of a loop amplitude is the one-loop MHV amplitude in the $\cN=4$ theory, 
which is given as sum over \lq\lq two-mass easy box functions\rq\rq  \cite{Bern:1993kr}.
Explicitly, one has  \cite{bddk}
\beq\label{fullampl}
A_{n;1}^{\cN=4 \, {\rm MHV}} =  A^{\rm
tree}_n\,  \sum_{i=1}^{n} \sum_{r=1}^{[{n\over 2}]-1}
\Bigl(
1 - {1\over 2} \delta_{{n\over 2} - 1, r}
\Bigr)\,
F_{n:r;i}^{2m\,e}
\  . 
\eeq
\begin{figure}[ht]
\begin{center}
\scalebox{0.65}{\includegraphics{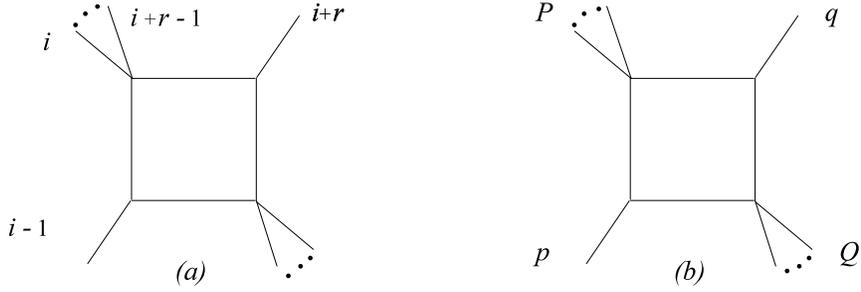}}
\end{center}
\caption{\it On the left we represent a two-mass easy box function $F_{n:r;i}^{2m\,e}$. $n$ is the total number of external legs, the labels $i$ and $r$ are defined in the figure. On the right we depict the same box function in a slightly simplified notation, used later in \eqref{boxcsw22}. The massless legs are called $p_{i-1} := p $ and $p_{i+r} := q$.  
  }
\label{Box_function}
\end{figure}
In order to define the two-mass easy box functions $F^{ 2me}$, we first introduce the  scalar box integral $I_4$, defined as  
\beq
I_4 = -i (4\pi)^{2-\epsilon}\, \int
\frac{d^{4-2\epsilon}p}{(2\pi)^{4-2\epsilon}}\,\,
\frac{1}{p^2(p-K_1)^2(p-K_1-K_2)^2(p+K_4)^2}
\ . 
 \eeq
In the two-mass easy case, two opposite momenta in the box integral are massless ($p$ and $q$ in Figure \ref{Box_function} (b)), and the remaining two opposite momenta  are massive ($P= k_i+\dots+k_{i+r-1}$ and $Q=k_{i+r+1}+\dots+k_{i+n-2}$ in the same figure). 
Denoting by $I_{4:r;i}^{2m\,e}$ the two-mass easy box integrals, the corresponding $F$ functions are defined  by 
\beqa\label{Ifunctions}
I_{4:r;i}^{2m\,e}  &=& - \frac{2F_{n:r;i}^{2m\,e}} {
              t^{[r+1]}_{i-1}  t^{[r+1]}_{i} -
              t^{[r]}_{i}  t^{[n-r-2]}_{i+r+1} }  \ , \nonumber \\
I_{4:i}^{1m}  &=&  - \frac{2F_{n:i}^{1m}} { t^{[2]}_{i-3}
                             t^{[2]}_{i-2} }
\ ,
\eeqa
where $
t_i^{[r]} = (k_i+\dots+k_{i+r-1})^2$
are the kinematic invariants constructed from sums of cyclically adjacent external momenta. Note that $P^2 = t^{[r]}_{i}$ and $Q^2=t^{[n-r-2]}_{i+r+1}$.
In the second equation in \eqref{Ifunctions} we have considered separately the one-mass case, i.e.~the special case where one of the two massive corners $P$ or $Q$ becomes massless. The corresponding function is obtained as a smooth limit of the two-mass easy configuration \cite{Bern:1993kr}.
Introducing the variables
\beq
\label{st} s \ := \ (P+p)^2 = t^{[r+1]}_{i-1} \ , 
\qquad   t \ := \ (P+q)^2 = t^{[r+1]}_{i} \ ,
\eeq
the explicit expression of the box function is \cite{bddk}
\beqa
\nonumber
F(s, t, P^2,
Q^2) & : = & -{1\over \epsilon^2} \Bigl[ (-s)^{-\epsilon} \, +\,
(-t)^{-\epsilon} \, -\, \big( (-P)^2\bigr)^{-\epsilon} \, -\,
\big( (-Q)^2\bigr)^{-\epsilon} \Bigr]
\\ \nonumber
&+& {\rm Li}_2 \Bigl( 1 - {P^2\over s } \Bigr) \, +\, {\rm Li}_2
\Bigl( 1 - {P^2\over t } \Bigr) +
{\rm Li}_2 \Bigl( 1 - {Q^2\over
s} \Bigr) \, + \, {\rm Li}_2 \Bigl( 1 - {Q^2\over t } \Bigr)
\\ 
&-& {\rm Li}_2 \Bigl( 1 - {P^2 Q^2\over s\, t } \Bigr)
\,
+\,
{1\over 2} \log^2 \Bigl( {s\over t} \Bigr)
\ , 
\label{boxcsw22}
\eeqa
where $P+p+Q+q=0$.
The relation to the functions $F_{n:r;i}^{2m\,e}$ is obtained by
setting $p=p_{i-1}$, $q=p_{i+r}$, and $P=p_i + \cdots +
p_{i+r-1}$.

Following the CSW proposal discussed above, one would immediately expect that the
one-loop MHV amplitude should arise from the one-loop MHV diagrams depicted in Figure \ref{mhv1loop}  below, i.e.~from gluing two MHV vertices together with two propagators in order to form a one-loop diagram, and summing all possible diagrams while preserving the cyclic ordering of the external legs.
\begin{figure} [ht]
\begin{center}
\scalebox{0.70}{ \includegraphics[width=4in]{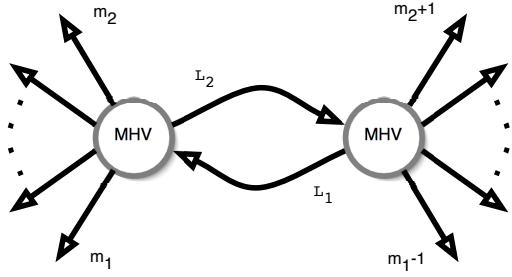}
}
\end{center}
\caption{\it One-loop MHV Feynman diagram,
using MHV amplitudes as interaction vertices,
with the CSW off-shell prescription.}
\label{mhv1loop}
\end{figure}
This calculation was performed  in  \cite{bst1}. 
One important point  to discuss  is the loop integral measure, 
\beq
\label{dem}
d\cM \ := \ {d^4 L_1 \over L_1^2 + i \varepsilon}
\,
{d^4 L_2 \over L_2^2 + i \varepsilon}
\
\delta^{(4)} (L_2 - L_1 + P_L)
\ ,
\eeq
where $L_1$ and $L_2$ are loop momenta, and $P_L$ is the
external momentum flowing outside the loop%
\footnote{In our conventions, all external momenta are
outgoing.}
so that $L_2 - L_1 + P_L=0$. Note that we also include the two propagators in the definition of the integration measure for convenience.

In order to calculate the one-loop MHV diagram in Figure \ref{mhv1loop},
 we need to re-express a  standard loop momentum 
in terms of the new variables $l$ and $z$
introduced in \eqref{offshell}. One finds that \cite{bst1}
\beq
\label{mmm}
 {d^4 L \over
L^2} \ = \ d\cN (l) \, {dz \over z} \ ,
\eeq
with the Nair measure  \cite{Nair}
\beq
 d\cN (l) := \, \lan l
\, \, dl \ran \, d^2 \tilde{l} \ - \ [ \tilde{l} \, \, d
\tilde{l}] \, d^2 l \ .
\eeq
Note that the product of the measure factor with 
a scalar propagator $d^4 L / L^2 $ of \eqref{mmm}
is independent of the reference vector $\eta$. Also,
the Lorentz invariant phase space measure for
a massless particle can be expressed  precisely
in terms of the Nair measure:
\beq
\label{nairmeas}
 d^4l \, \delta^{(+)} (l^2)\ =  \
{d\cN (l)\over 4i} \ ,
\eeq
where, as before, we write the null vector $l$ as $l_{\a \da} =
l_{\a} \tilde{l}_{\da}$, and in Minkowski space we identify
$\tilde{l} = \pm l^{\ast}$ depending on the sign of the energy. 

We are now in position to write the one-loop integration measure  in terms of variables that are appropriate for the MHV diagram method. 
Expressing  $L_1$ and $L_2$ as in
\eqref{offshell},
\beq
\label{ells}
 L_{i; \a, \da} \ =
\ l_{i \a} \llt_{i \da}  \, + \,
z_i \, \eta_{\a}\tilde{\eta}_{\da} \ ,
\qquad i=1,2 \ , 
 \eeq
we  rewrite the argument of the delta function in \eqref{dem}  as
\beq
L_2 - L_1 + P_L
= l_2 - l_1 + P_{L; z} \ ,
\eeq
where
$P_{L;z} := P_L - z \eta$, 
and
$z \ := \ z_1\, - \, z_2$. 
Notice that we use the same $\eta$ for both the momenta
$L_1$ and $L_2$.
Using \eqref{ells}, we can re-cast \eqref{dem} as
\cite{bst1}
\beq
\label{gcar}
d\cM \ = \
{dz_1 \over z_1 + i \vare_1}\, { dz_2 \over z_2 + i \vare_2}
\
\bigg[ {d^3 l_1 \over 2 l_{10}}\,
{d^3 l_2 \over 2 l_{20}}
\,
\delta^{(4)}  (l_2 \, - \, l_1 \,  + \, P_{L; z} )
\bigg]
\ ,
\eeq
where
$\vare_{i} := {\rm sgn}(\eta_0 l_{i0}) \vare =
{\rm sgn}(l_{i0})\vare$, $i = 1, 2$ (the last equality
holds since we are assuming $\eta_0 >0$).

The next step involves integrating out the variable $z':= z_1 + z_2$, which only appears  through  the integration measure 
\beq
{dz_1 \over (z_1 + i \vare_1)}\, { dz_2 \over   (z_2 + i \vare_2)} := 2\, {dz \, dz'  \over \big[  (z' + z + i \vare_1)(z' - z + i \vare_2)\big]}
\  .
\eeq  
This can be done using the residue theorem, with the result \cite{bst1, ftt} that one can simply replace 
\beq
{dz_1 \over (z_1 + i \vare_1)}\, { dz_2 \over   (z_2 + i \vare_2)} \ \rightarrow \
 2 \pi i \, {dz \over z + i \vare}
\ .
\eeq
The last crucial step consists in performing the integration over $z$. Perfectly in tune with the S-matrix programme, 
this was converted in \cite{bst1} into a dispersion integral. The final result for the integration is then \cite{bst1,ftt}
\beq
\label{bstmeasure}
d\cM \ = \
 2 \pi i \
\theta (P_{L;z}^2)\
{dP_{L;z}^2 \over P_{L;z}^2 \, - \, P_L^2 \, - \, i \vare}
\
d{\rm LIPS} (l_2^{\mp} , - l_1^{\pm}; P_{L;z})
\ , 
\eeq
where
\beq
\label{LIPS}
d{\rm LIPS} (l_2^{-}  , - l_1^{+} ; P_{L;z}) \ := \
d^4 l_1 \, \delta^{(+)} (l_1^2) \
d^4 l_2 \, \delta^{(-)} (l_2^2 )\
\delta^{(4)} (l_2 - l_1 + P_{L;z})
\
\eeq
is the two-particle Lorentz invariant phase
space  (LIPS) measure, and we
recall that $ \delta^{\pm} (l^2) : =
\theta ( \pm l_0 ) \delta (l^2)$.   

It is important to observe that the integration is performed for
$P_{L;z}^2 > 0$. By setting all the various external
kinematical invariants $P_{L}^2$ to negative values,
no poles are encountered along the integration contour
and the $i \vare$ prescription can be dropped.
However, \eqref{bstmeasure} provides
us with the correct analytic continuation to
the physical region, which is obtained by simply performing
the substitution
$
P^2_L \ \longrightarrow \ P^2_L \, + \, i \vare
$. 

Having fully re-expressed the integration in terms of MHV variables, one can now  perform all the remaining integrations and sum over all possible MHV diagrams corresponding to the different choices of
particles running in the internal legs on each of the two MHV vertices. 
The interested reader can consult \cite{bst1} for the remaining part of the calculation; in the following we will outline its  main steps.

 Firstly, one finds that each MHV  diagram leads to a sum of four terms which are written as dispersion integrals in the channel fixed by the diagram. These are related to unitarity cuts of four different box functions. Crucially, in order to reconstruct complete box functions one has to take an appropriate combination of four dispersion integrals from four different MHV diagrams (channels).
Surprisingly, it was found in \cite{bst1} that dependence on the reference spinor $\eta$ cancels out between the four dispersion integrals contributing to a single box function and not only in the sum over all diagrams.
The covariance at one-loop for arbitrary amplitudes was later shown in \cite{ftt} using the Feynman tree theorem. Collecting all the box functions, one finds precisely the expected result \eqref{fullampl}.

We would like to make a few comments before closing this subsection. 

Firstly, the CSW approach, and hence the general MHV diagram proposal for calculating scattering amplitudes, was originally motivated by the twistor space localisation of amplitudes and
Witten's formulation of $\cN=4$ super Yang-Mills theory as a theory in twistor space.
At first sight, then, one might not expect  these methods to work for theories with less supersymmetry (at tree level of course the gluon scattering amplitudes are the same for the supersymmetric and non-supersymmetric cases).  Nevertheless, one could just put aside considerations of the twistor picture, and simply see if the MHV diagrams do yield the correct
amplitudes in Yang-Mills theories with less than maximal supersymmetry. This was done in \cite{quig, bbst1}, where results from the MHV diagram calculation in $\cN=2$ and $\cN=1$ super Yang-Mills were found in agreement with those derived in \cite{fusing} using unitarity.  This success was somewhat unexpected, and  
led to a new, simpler understanding of the twistor space localisation of loop amplitudes  in these theories \cite{csw2, csw3}. 
We also point out that this success was not accidental. Indeed, in \cite{ftt} it was proved that any non-MHV amplitude in supersymmetric gauge theories is reproduced correctly by MHV diagrams.

Finally, the 
non-supersymmetric Yang-Mills theory  
was studied at one loop using MHV methods \cite{bbst2}. Again, this correctly yielded the (cut-constructible part of the) amplitude,
and this work was notable in that it included the first calculation of a previously unknown amplitude using these methods.
It was at this point that the issue of so-called rational terms arose -- these are terms in amplitudes
that are not generated by unitarity cuts performed in four dimensions. This problem is not present in supersymmetric Yang-Mills theories, but does appear in the non-supersymmetric theory. The MHV diagram approach, as used in
\cite{bbst2}, reproduced correctly all the non-rational terms, but not the rational ones.
Note however that the MHV vertices that one inserts into the diagrams are essentially four-dimensional objects, although one does use dimensional regularisation in the loop integral; 
it is expected that it is this fact that underlies the missing rational pieces, although this remains
to be investigated fully. Calculation of the rational terms has been achieved by various methods that are detailed e.g. in the
chapter of this review from Britto \cite{britto}.

\subsection{Supersymmetric MHV diagrams} 
It is generally the case that formalisms where symmetries are explicit are more powerful both conceptually and
for calculations. It is easy to generalise MHV diagrams in $\cN=4$ SYM to a manifestly  supersymmetric version. 
This is particularly convenient when one is interested in amplitudes with external fermions or scalars. The on-shell superamplitude formalism introduced in the following is also reviewed in the chapter by Elvang, Freedman and Kiermaier \cite{efk}.

The $\cN=4$  theory contains 
two gluons $G^{\pm} (p)$ with helicities $1, -1$,  four fermions $\psi_A$ of helicity $+1/2$,  transforming in the fundamental of the R-symmetry group $SU(4)_{\mathrm{R}}$ of the $\cN=4$ theory,  four Weyl fermions $\bar\psi^A$ with helicity $-1/2$  in the anti-fundamental representation,  and six real scalar fields (corresponding to particles of zero helicity)  $\phi_{[AB]}$ in the antisymmetric tensor representation of the R-symmetry group. Here $A, B=1,\ldots , 4$ are fundamental $SU(4)_{\mathrm{R}}$ indices. 

Following Nair, one introduces four auxiliary Grassmann variables%
\footnote{Not to be confused with the reference spinor of the same name introduced in Section \ref{section-offshell}.}
$\eta^A$,  and combines all the physical states of the theory into a super-wavefunction, 
\beq 
\label{superNair} 
\Phi (\eta, p) \, := \, G^+(p) + \eta^A \psi_A (p)  + {\eta^A \eta^B \over 2!} \phi_{[AB] } (p) + 
\eps_{ABCD}  {\eta^A \eta^B \eta^C\over 3!}  \bar{\psi}^D(p)  + \eta^1\eta^2 \eta^3 \eta^4
 G^{-} (p) 
 \ . 
 \eeq
All amplitudes with  a fixed total helicity are then collected into a single object called a superamplitude. The supersymmetry charges $q^A$, $\bar{q}_A$ satisfy the algebra $\{ q^A_a, \bar{q}_{B \dot{a}} \} = p_{a \dot{a}} \delta^A_B$ where, on shell, $p_{a \dot{a}}  := \lambda_a \lt_{\dot a}$. In terms of the 
$\eta$'s these are realised as 
\beq
q^A _a \ = \ \sum_{i=1}^n \lambda_{ia} \eta^A_i
\ , 
\qquad 
\bar{q}_{A \dot{a}}  \ = \ 
\sum_{i=1}^n \lt_{i\dot{a}} {\del \over \del \eta^A_i} 
\ . 
\eeq
One can realise the $q$-supersymmetry manifestly on the amplitudes by pulling out a $\d$-function of supermomentum conservation, $ \delta^{(8)} (\sum_{i=1}^n \l_i \eta_i )$. 
The MHV  superamplitude is then  expressed  as \cite{Nair}   
\beq 
\label{superMHV} 
\cA_{n, {\rm MHV} }(1, \ldots , n) \, := \, i \delta^{(8)} (\sum_{i=1}^n \l_i \lt_i ) \delta^{(4)}  (\sum_{i=1}^n \l_i \eta_i )\prod_{i=1}^n {1\over \lan i i+1 \ran}
\ , 
\eeq
with 
$\l_{n+1} \equiv \l_1$. 
As one can see from \eqref{superNair}, in order to scatter a particular external state with  
helicity $h_i$, we need to expand (\ref{superMHV}) and pick the term containing $p_i = 2 - 2h_i$ powers of  $\eta_i$. 
For example, the gluon MHV amplitude (\ref{MHV}) with negative helicity gluons $i^-$ and $j^-$ arises as  the coefficient of $\eta_i^4 \eta_j^4$ in the expansion of 
\eqref{superMHV}.
 
As an example, we will now apply this formalism to derive a compact formula for all
NMHV tree-level amplitudes in $\cN=4$ super Yang-Mills. In this case we have to connect two MHV superamplitudes with an appropriate off-shell continuation of the internal leg $P$. The off-shell continuation for the
spinor variables is as usual $\lambda_{P;a} = P_{a \dot{a}} \xi^{\dot{a}}=
-i \lambda_{-P;a}$ where
$\xi$ is the reference spinor. Both MHV supervertices have degree 8 in $\eta$ while the NMHV superamplitude has degree 12. This suggests that we
have to augment the super-MHV rules with an $\int\!\!d^4\eta_P$ integration for every internal leg $P$ 
where $\eta_P = -i \eta_{-P}$ are the fermionic variables for the internal legs. 
It is easy to see that in terms of component amplitudes this guarantees that the correct helicity state propagates between the two MHV vertices while it reduces the total Grassmann degree to $16-4=12$ as expected.

We now move on to the calculation. There is only one type of diagram, depicted in  Figure \ref{supernmhv} below. Note that we do not assign specific helicities to external legs as we are considering superamplitudes and that for a superamplitude only the total helicity $\sum_{i=1}^n h_i$ is fixed.
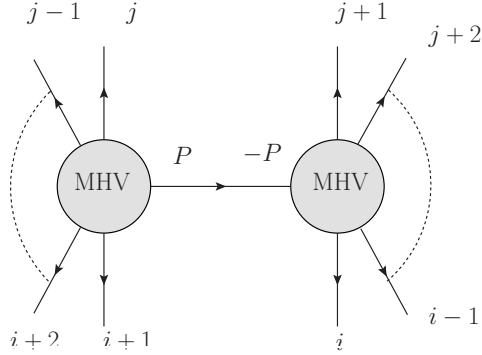
\begin{figure}[h]
\label{supermhv}
\begin{center}
\scalebox{0.55}{
\fcolorbox{white}{white}{
  \begin{picture}(324,240) (95,-59)
    \SetWidth{0.8}
    \SetColor{Black}
    \GOval(160,48)(32,32)(0){0.882}
    \GOval(320,48)(32,32)(0){0.882}
    \Line[arrow,arrowpos=0.5,arrowlength=5,arrowwidth=2,arrowinset=0.2](160,16)(160,-48)
    \Line[arrow,arrowpos=0.5,arrowlength=5,arrowwidth=2,arrowinset=0.2](160,80)(160,144)
    \Line[arrow,arrowpos=0.5,arrowlength=5,arrowwidth=2,arrowinset=0.2](144,76)(112,135)
    \Line[arrow,arrowpos=0.5,arrowlength=5,arrowwidth=2,arrowinset=0.2](144,20)(112,-39)
    \Line[arrow,arrowpos=0.5,arrowlength=5,arrowwidth=2,arrowinset=0.2](192,48)(288,48)
    \Line[arrow,arrowpos=0.5,arrowlength=5,arrowwidth=2,arrowinset=0.2](320,80)(320,144)
    \Line[arrow,arrowpos=0.5,arrowlength=5,arrowwidth=2,arrowinset=0.2](334,77)(367,136)
    \Line[arrow,arrowpos=0.5,arrowlength=5,arrowwidth=2,arrowinset=0.2](336,19)(368,-40)
    \Line[arrow,arrowpos=0.5,arrowlength=5,arrowwidth=2,arrowinset=0.2](320,16)(320,-48)
    \Arc[dash,dashsize=2,clock](296.541,48.942)(87.464,48.058,-48.933)
    \Arc[dash,dashsize=2](188.972,47.489)(92.973,135.201,223.918)
    \Text(176,160)[lb]{\Large{\Black{$j$}}}
    \Text(320,160)[lb]{\Large{\Black{$j+1$}}}
    \Text(384,144)[lb]{\Large{\Black{$j+2$}}}
    \Text(384,-48)[lb]{\Large{\Black{$i-1$}}}
    \Text(320,-64)[lb]{\Large{\Black{$i$}}}
    \Text(160,-64)[lb]{\Large{\Black{$i+1$}}}
    \Text(96,-64)[lb]{\Large{\Black{$i+2$}}}
    \Text(108,160)[lb]{\Large{\Black{$j-1$}}}
    \Text(140,46)[lb]{\Large{\Black{$\rm MHV$}}}
    \Text(304,46)[lb]{\Large{\Black{$\rm MHV$}}}
    \Text(256,64)[lb]{\Large{\Black{$-P$}}}
    \Text(208,64)[lb]{\Large{\Black{$P$}}}
  \end{picture}
}}
\end{center}
\caption{\it The super-MHV diagrams contributing to the NMHV superamplitude.}
\label{supernmhv}
\end{figure}
Evaluating this, we are directly led to
\beqa\label{supernmhvamp}
\mathcal{A}_{n,{\rm NMHV}} \ &\!\!\!\!\! = & \sum_{i,j} 
\int d^4\eta_P 
\mathcal{A}_{\rm MHV}(i+1 , \ldots , j)   \frac{i}{P^2} \mathcal{A}_{\rm MHV} (j+1, \ldots , i)  \nonumber\\
&\!\!\!\!\!=& i^3 \sum_{i,j} \frac{1}{\prod_{k=1}^n \langle k k+1 \rangle} \frac{\langle i i+1 \rangle \langle j j+1 \rangle}{\langle j P \rangle \langle P i+1 \rangle
\langle i -\!\!P \rangle \langle -\!P j+1 \rangle} \frac{1}{P^2} \times {\mathcal F}
\, , 
\eeqa
where the sum $\sum_{i,j}$ is taken over all inequivalent MHV superdiagrams, and
\beqa
{\mathcal F}  & = &  \int d^4\eta_P \, \delta^{(8)}(\lambda_{Pa} \eta_P^A + q_{a,{\rm left}}^A) \delta^{(8)}(\lambda_{-Pa} \eta_{-P}^A + q_{a,{\rm right}}^A) \nonumber \\ 
& = &  \delta^{(8)}(q_{a,{\rm left}}^A + q_{a,{\rm right}}^A) \int d^4\eta_P \delta^{(8)}(\lambda_{Pa} \eta_P^A + q_{a,{\rm left}}^A) \nonumber \\
& = & \delta^{(8)}(q_{a,{\rm tot}}^A) \prod_{A=1}^4 (\sum_{k=i+1}^j \langle P k \rangle \eta_k^A)\ , 
\eeqa
with $q_{\rm left} = \sum_{k=i+1}^j  \lambda_k \eta_k$, 
$q_{\rm right} = \sum_{k=j+1}^i  \lambda_k \eta_k$, $q_{\rm tot} = q_{\rm left}+q_{\rm right}$, and $P=k_{j+1} + \cdots + k_i$. Note that  in the last line we factored out an overall supermomentum conservation delta function and performed the $\eta_P$ integrations. Hence, we find
\beq
\mathcal{A}_{n,{\rm NMHV}}\ =\ \mathcal{A}_{n,{\rm MHV}}\,  
\frac{\langle i i+1 \rangle \langle j j+1 \rangle}{\langle j P \rangle \langle P i+1 \rangle
\langle i P \rangle \langle P j+1 \rangle} 
\, 
\frac{1}{P^2}\, 
\prod_{A=1}^4 \Big(\sum_{k=i+1}^j \langle P k \rangle \eta_k^A\Big) \, ,
\eeq
where we have factored out the $n$-point MHV superamplitude \eqref{superMHV}.

 \subsection{Lagrangian derivation of MHV diagrams}
 \label{cov}
 
We have discussed above how MHV diagrams correctly yield amplitudes in Yang-Mills theories. 
If one prefers the view of the \lq\lq constructive S-matrix\rq\rq\ approach, which uses primarily analytic properties and on-shell quantities as input to calculate the scattering matrix, then one might conclude that the methods discussed above, and in the following section, are perfectly adequate. In fact it turns out that they are superior in many ways to traditional approaches. Yet it is natural to ask if one can {\it derive} the MHV rules  directly from a Lagrangian --  essentially as some sort of Feynman diagrams. This might also have the advantage of illuminating how the quantum theory of MHV diagrams is best formulated in general, and also assist with exploring possibly interesting further structures and different kinds of perturbative expansions of the S-matrix.
 
Such a Lagrangian description has indeed been formulated, as we will now describe. The key steps were taken in
\cite{Gorsky:2005sf,Mansfield:2005yd}, and subsequent work can be found in  
\cite{Ettle:2006bw,Ettle:2007qc, Feng:2006yy, Brandhuber:2006bf, Brandhuber:2007vm}.
 A more recent article \cite{Fu:2009nh} may also be consulted for further references; in the following summary we will use the notation of  this paper. 

The starting point is to quantise  the Yang-Mills theory in the lightcone gauge, and
 investigate whether a suitable canonical transformation can take one from the lightcone Yang-Mills Lagrangian to one whose terms precisely lead to the MHV vertices. We observe that such a MHV Lagrangian must therefore have an infinite number of terms. 
If one chooses the Minkowski space lightcone coordinates
 \beq\label{lccoords} 
  \hat x = \frac{1}{\sqrt 2}(t-x^3),
\quad \check x = \frac{1}{\sqrt 2}(t+x^3), \quad z = \frac{1}{\sqrt
2}(x^1+ix^2), \quad \bar z = \frac{1}{\sqrt 2}(x^1-ix^2),
\eeq
with gauge-fixing condition $\hat
A=0$, then, after elimination of unphysical degrees of freedom,  the Yang-Mills action can be written in terms of positive-
and negative-helicity fields ${\mA}\equiv A_z$ and ${\bar \mA}\equiv
A_{\bar z}$ as
\beq
\label{mansfieldaction}
S = \frac 4{g^2} \int d\hat x\:\int_\Sigma d^3{\bf x}\:({\cal L}^{-+}+{\cal L}^{-++}+{\cal L}^{--+}+{\cal L}^{'--++}),
\end{equation}
with
\beqa
\label{mansfieldterms}
{\cal L}^{-+} &= \phantom{-}{\rm Tr}\,
{\bar \mA}\,\left(\check\partial\hat\partial-
\partial\bar\partial\right)\,\mA\,, \nonumber \\
{\cal L}^{-++}&=-{\rm Tr}\,
({\bar\partial}{\hat\partial}^{-1} {  \mA})\:
[{  \mA},{\hat\partial} {\bar\mA}]\,,  \nonumber \\
{\cal L}^{--+}&=-{\rm Tr}\, [{\bar
{\cal A}},{\hat\partial} {  \mA}]\:
({  \partial}{\hat\partial}^{-1} {\bar \mA})\,,  \nonumber \\
{\cal L}^{'--++}&=-{\rm Tr}\, [{\bar \mA
},{\hat\partial} { \mA }]\:{\hat\partial}^{-2}\: [{ \mA
},{\hat\partial} {\bar \mA }]\,,
\eeqa
with $\Sigma$ is a constant-$\hat x$ quantisation surface and $d^3{\bf
x}=d\check x\,dz\,d\bar z$.

The first two terms above, ${\cal L}^{-+} +{\cal L}^{-++}$,  describe self-dual Yang-Mills theory
 \cite{Chalmers:1996rq}. This is a free theory at tree level, and the only non-vanishing amplitude is at one loop, when all particles have positive helicity. Classically, therefore, one would expect there to be a canonical change of variables which transforms these two terms into a free Lagrangian. Formally, we seek  a new field
$\mB$, a (non-local) functional of $\mA$ on the surface of
constant $\hat x$, such that ${\cal L}^{-+}+ {\cal L}^{-++}$ can be
written as a free theory,
\beq
\label{eq:transform-4d}
{\cal L}^{-+}[\mA, {\bar\mA}] + {\cal L}^{-++}[\mA, {\bar\mA}] = {\cal L}^{-+}[\mB,{\bar\mB}]\,,
\eeq
with the condition on ${\bar\mB}$ that the transformation be canonical:
\beqa
 \label{canontr} 
 &&\hat\partial \bar\mA^a(\hat x,\vec x) = \int_\Sigma d^{3}\vec y \:
 \frac{\delta\mB^b(  \hat x,\vec y)}{\delta\mA^a(\hat x  ,\vec x)} \hat\partial
 \bar\mB^b(\hat x  ,\vec y)  \nonumber\\ &&
 \Leftrightarrow \hat\partial
\bar\mB^a(\hat x,\vec x) = \int_\Sigma d^{3}\vec y \:  \frac{\delta\mA^b(  \hat x,\vec y)}{\delta\mB^a(\hat x  ,\vec x)} \hat\partial  \bar\mA^b(\hat x  ,\vec y).
\eeqa
After Fourier transforming, one can show that $\mB$ is a power series in $\mA$,  
\beqa
\label{BtoA}
&&\mB(\hat x,\vec p)=\mA(\hat x,\vec p)+
\sum_{n=2}^\infty \int {d^3k_1\over (2\pi)^3}\dots{d^3k_n\over (2\pi)^3} 
 \times \nonumber \\
&&
{{\hat p}^{n-1}\,(2\pi)^3\,\delta^3 (\vec p-\sum\vec k_i)\over
(p,k_1)\,(p,k_1+k_2)\dots(p,k_1+\dots+k_{n-1})}  
\mA(\hat x,\vec k_1)\dots
 \mA(\hat x,\vec k_n) \ ,
 \nonumber\\ 
\eeqa
where we have introduced $(i,j) := \hat{i} \tilde{j} - \hat{j} \tilde{i} = p^i_+ p^j_z - p^j_+ p^i_z$. Furthermore, we note that $\bar\mA$ is a power series in $\bar\mB$, which is also linear in $\mB$,
\beqa
\label{AbartoB}
&&\bar\mA(\hat x,\vec p)=\bar\mB(\hat x,\vec p)+\nonumber\\
&&
\sum_{m=3}^\infty \sum_{s=2}^m\int {d^3k_1\over (2\pi)^3}\dots{d^3k_n\over (2\pi)^3} {{\hat k_s}\over \hat p}\,\Xi^{s-1}(\vec p,
-\vec k_1,\dots,-\vec k_m)\times\nonumber\\
&&
(2\pi)^3 
\,\delta^3 (\vec p-\sum\vec k_i)  \mB(\hat x,\vec
k_1)\dots\bar\mB(\hat x,\vec k_s)\dots
 \mB(\hat x,\vec k_m)\ ,
\eeqa
for certain known coefficients $\Xi^{s-1}(\vec p,
-\vec k_1,\dots,-\vec k_m)$.   One can then see that when the lightcone action is written in terms of the new variables $(\mB,\bar\mB)$, there is the free, quadratic term plus an infinite sequence of interaction terms, each of which
has precisely two fields $\mB$, together with increasing numbers of fields $\bar\mB$, 
\beq
\label{BLagrangian}
 {\cal L}[\mA, {\bar\mA}]={\cal L}^{-+}[\mB,\bar\mB] + {\cal L}^{--+}[\mB,\bar\mB ] + {\cal L}^{--++}[\mB,\bar\mB ] + {\cal L} ^{--+++}[\mB,\bar\mB ] + \cdots.
\eeq
Explicit calculation  \cite{Ettle:2006bw} shows that
these interaction terms precisely give the Parke-Taylor amplitudes, continued off-shell
in exactly the same way as proposed by \cite{csw1}. Note that the reference null-momentum $\eta^{a \dot{a}} = \eta^a \eta^{\dot{a}}$ thus comes to be identified with the gauge fixing vector in the lightcone action.

We thus seem to have a candidate Lagrangian for generating the MHV rules. However,
some care is needed here -- to start with, the canonical transformation that we have used to change variables is non-local. This may affect the S-matrix equivalence theorem, by which we 
would like to conclude that the $(\mB,\bar\mB)$ variables may equivalently be used to describe the quantum theory. 
This theorem may in principle be violated by
$1/p^2$ terms arising from the non-locality   -- in short, these could cancel LSZ reduction factors of $p^2$ and hence survive in the on-shell limit.
In addition, and partly related to this, the Lagrangian above clearly cannot directly generate all amplitudes -- 
e.g.~the non-zero one-loop all-plus helicity amplitudes clearly cannot be obtained from the MHV rules.

A more careful analysis reveals the following. Firstly, it has been found that the only diagrams
that violate the equivalence theorem are certain dressed tadpole diagrams  \cite{Fu:2009nh} that explain e.g. the presence of the rational
all-plus amplitudes, and apart from these one may freely use the $(\mB,\bar\mB)$ variables to calculate amplitudes. As far as the rational parts of other amplitudes are concerned,
note that the transformation of variables above can be defined in $D=4-2\epsilon$ dimensions and the resulting MHV vertices are also $D$ dimensional -- this differs from the MHV diagram method described earlier where four-dimensional vertices are used. The regularisation employed may be the
four-dimensional helicity  scheme \cite{Bern:1991aq}, where the internal momenta are in $D$ dimensions whilst the external particles carry four-dimensional helicities, as in 
 \cite{Ettle:2006bw}, or standard dimensional regularisation as in \cite{Ettle:2007qc}.
 In either case, the additional $\epsilon$-dependent terms arising from these vertices are expected to combine with divergent terms from loop integrations to produce the missing rational terms of the amplitudes, although this remains to be investigated fully.
 
 Another way to produce the rational terms has been described in \cite{Brandhuber:2007vm}.
 This is based on the lightcone gauge formulation of Yang-Mills theory given in 
\cite{Thorn05, CQT1, CQT2}, which uses a regularisation in four dimensions most naturally formulated in terms of region 
(or T-dual) momenta.  
In this approach, a very simple two-point counterterm reproduces all the $n$-point all-plus amplitudes. It appears plausible that all
other rational parts of amplitudes will arise in a similar way, although this has not been shown explicitly.

In conclusion, a Lagrangian containing interaction terms that reproduce the MHV vertices has been shown to be derived from ordinary Yang-Mills theory.
However, in general there is a rather subtle interplay of (violations of) the S-matrix equivalence theorem and the regularisation scheme used, which is responsible for generating the rational parts of amplitudes. Whilst the schemes mentioned above do correctly generate these terms
in the cases studied, the calculations involved appear to become difficult in full generality.
  
We also mention that an alternative derivation of the MHV diagram method directly in twistor space was found in \cite{Boels:2007qn}  where, 
building upon \cite{lion, Boels:2006ir}, the  MHV diagram expansion was derived  from a particular axial gauge fixing of a supersymmetric twistor action. See also \cite{adamason} for very recent related work.

\subsection{Other directions} 
  
To conclude this section, we would like to mention some other important applications of, or contributions to MHV diagrams.   
  
{\bf 1.} Recent work \cite{Bullimore:2010pj} (see also \cite{Mason:2010yk,Brandhuber:2010mi}) has  
described a momentum twistor space \cite{momhodges} formulation of the MHV diagram approach which is, roughly speaking, 
dual to that described above, and in which dual superconformal quantities play a key role.
In this approach, the MHV vertices are replaced by unity, whilst the internal propagators are
represented by superconformal $R$-functions well known from other work. In this approach,
features such as symmetries are readily apparent, and one may relatively easily write
down integrands for quite general amplitudes and study their properties. The relationship with
Wilson loops in twistor space can also be explored.  A discussion of the space-time analogues of this work has been given recently in \cite{Brandhuber:2010mi}.

{\bf 2.} 
In an interesting paper \cite{dgk}, Dixon, Glover and Khoze managed to apply the MHV diagram method to the effective Lagrangian describing the one-loop top quark contribution to Higgs plus multigluon scattering processes, in the heavy top limit. In this limit, such processes are conveniently described by a dimension five operator of the form $H \, {\rm Tr}\, (F_{\mu \nu}F^{\mu \nu})$, where $H$ is the Higgs field.  Instead of discussing directly the amplitudes arising from that operator, the authors of \cite{dgk} considered the interaction  
$\phi \,  {\rm Tr}\, F^2_{\rm SD}\, + \, \bar\phi  \, {\rm Tr}\, F^2_{\rm ASD}$, where $\phi$ is a complex field whose real part is equal to $H$, $H := \phi + \bar\phi$, and  
$F_{\rm(A)SD}$  stands for (anti)self-dual part of the field strength. Because of the last equation, the Higgs amplitudes can be recovered by summing those with one $\phi$ and those  with one $\bar \phi$. The key point is that the $\phi$ and $\bar\phi$ amplitudes are amenable to a derivation through MHV rules, which makes their calculation much more efficient.  
This approach was extended to one loop for the same amplitudes in \cite{BGR}.
Further extensions of the MHV diagrammatic method include applications to QED \cite{Ozeren} and Yang-Mills with coloured massive scalars \cite{BoelsSchwinn1, BoelsSchwinn2}.

\section{BCFW recursion relation}

As we have alluded to above, one of the main ideas in the analytic S-matrix programme is that of calculating scattering amplitudes 
from the knowledge of their analyticity properties, 
possibly without even knowing the Lagrangian of the theory \cite{S}.  
A second key idea is to study amplitudes as functions of {\it complex variables}.
We note here in passing that also for twistor theory the most natural arena is complexified Minkowski space. 
The BCFW recursion relation incorporates neatly these two precepts. It provides us with 
an algorithm to calculate efficiently, and in a recursive way,  all tree-level scattering amplitudes for various theories satisfying certain prerequisites (to be reviewed shortly),  based on the knowledge of the singularities of amplitudes. At tree level, these are  simple poles  in the two-particle and multi-particle kinematic invariants, associated to collinear and multiparticle singularities, respectively.
Starting from the smallest building blocks, namely three-point amplitudes, all amplitudes can then be constructed recursively. This is a very  powerful statement --  for example this implies that one can reconstruct the S-matrix of General Relativity  simply from the knowledge of the three-point graviton amplitudes. 
We remark that these are just the squares of the three-point Yang-Mills amplitudes, which is the basis for another
very interesting line of enquiry, relating gravity amplitudes to Yang-Mills ones. This is reported elsewhere in this review. 
In order to discuss the recursion relations, we consider first the case of colour-ordered amplitudes in massless Yang-Mills theory, and later generalise to other situations. 

A key feature of the BCFW recursion relation is that of mapping a subset of simple poles  of the tree amplitude into simple poles in a single auxiliary complex variable $z$, whose residues are then calculated from well-known factorisation properties of amplitudes. 
In its simplest incarnation, one proceeds as follows. 

\subsection{Derivation of the recursion}
Consider a generic amplitude $A(p_1, \ldots, p_n)$, and select two legs for special treatment; without loss of generality we can choose these to be 1 and 2 (note that we can also shift non-adjacent particles but this would lead to recursion relations involving more terms).
One then shifts the two  momenta as 
\beq
\label{shift}
\hat{p}_1 (z) = p_1 + z \eta \ , \qquad \hat{p}_2 (z) = p_2 - z \eta 
\ . 
\eeq
The shifts performed in \eqref{shift} are chosen in a particular form in order not to alter the momentum conservation condition. Furthermore, we would like to preserve the on-shell condition for particles 1 and 2, which is possible if $p_1 \eta = p_2 \eta = 0$. In real Minkowski space there are no solutions to these constraints but in complex Minkowski space there are two solutions, 
$\eta= \l_1 \lt_2$ and $\eta = \l_2 \lt_1$, where $p_{i} = \l_{i} \lt_{i}$, $i=1,2$,  as usual. Picking for example the first solution, we rewrite \eqref{shift} as 
\beq
\label{bcfshifts}
\hat{p}_1 (z) = \l_1 ( \lt_1 + z \lt_2) := \l_1 \hat{\lt}_1 \ , 
\ \quad
\hat{p}_2 (z) = (\l_2 - z \l_1)  \lt_2 := \hat{\l}_2 \lt_2 \ . 
\eeq
We can define the complex function 
$A (z) := A (\hat{p}_1, \hat{p}_2 , p_3 , \ldots , p_n)  $. 
Since $\hat{p}_1^2 (z) = \hat{p}_2^2 (z)= 0$ for all values of $z$, $A (z)$ defines a one-parameter family of scattering amplitudes. Importantly, the shifts introduced earlier make the kinematic variables
complex, hence we should regard $A(z)$ as describing a scattering process in complexified Minkowski space. 

It is very easy to  prove that $A(z)$ is a rational function of $z$, which has only simple poles in this variable. 
Indeed, singularities in tree amplitude  arise from a Feynman propagator which is going on shell.  Let $\hat{P}_{ij} = p_i + \cdots p_j$ be the momentum of a certain propagator. There are three possibilities: either leg one or two belong to $\hat{P}_{ij}$; or, either both legs, or none, belong to $\hat{P}_{ij}$. It is only in the first case that $\hat{P}_{ij}$ depends on $z$, since in the other two remaining cases, this dependence is either not present or it cancels since $\hat{p}_{1} + \hat{p}_{2} = p_{1} + p_{2}$.
Focusing on the first case, and assuming for definiteness that particle 1 belongs to $ \hat{P}_{ij}  $, we can write 
$\hat{P}_{ij} = P_{ij} - z \l_1 \lt_2$, and 
\beq
\label{ppp}
{1\over \hat{P}_{ij}^2} \ = \ - {z_{ij}\over P_{ij}^2} \, {1\over z - z_{ij} } \ , 
\eeq
where $z_{ij}$ is the solution of $\hat{P}_{ij}^2 = 0$, namely 
\be
\label{polesinz}
z_{ij} = {P_{ij}^2  \over  \lan 1 | P_{ij} | 2]}
\ . 
\ee 
All the poles in $z$ of $A(z)$ are given by \eqref{polesinz}.  
Importantly, we also know the residues at these poles. Indeed, since $A(z)$ is a physical amplitude, the residue will be given by factorisation on the corresponding pole. Specifically, as $z\to z_{ij}$, we have 
\beq
\label{h1}
A(z) \to \sum_{h}   A_{L}^{h} (z= z_{ij}) {i\over \hat{P}_{ij}^2(z)}  A_{L}^{-h} (z= z_{ij}) \ , 
\eeq
hence 
\be
\label{h2}
{\rm Res} \left.A(z)\right|_{z = z_{ij}} \ = - \sum_{h}A_{L}^{h} (z= z_{ij}) \, i {z_{ij}\over P_{ij}^2}  \, A_{L}^{-h} (z= z_{ij}) \ , 
\eeq
where we have used \eqref{ppp}. In \eqref{h1} and \eqref{h2} we have included a sum over all possible internal helicities, as required from factorisation theorems. 

\begin{figure}[ht]
\begin{center}
\scalebox{0.55}{\includegraphics{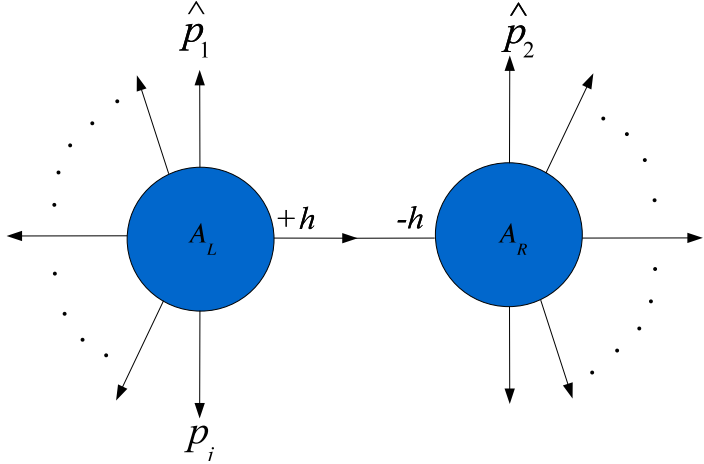}}
\end{center}
\caption{\it One of the recursive diagrams contributing to the BCFW recursion relation for a colour-ordered amplitude $A(1, \ldots , n)$. The particles with shifted momenta are adjacent -- namely 1 and 2. Therefore, there is a single sum in the recursion relation, labeled by $j$.    }
\label{rec_final_amplitude}
\end{figure}

One can therefore consider the following integral over a large circle at infinity $C_{\infty}$,  
\beq
\label{sop} 
A_{\infty}:= {1\over 2 \pi i} \oint_{C_\infty}\!{dz \over z} \, A(z) \ = \ A(0) \, + \,  \sum_{z_{ij}\neq 0} {\rm Res} {A(z)\over z} 
\ . 
\eeq
Here we have defined $A_{\infty} := A( z\to \infty)$. Clearly it is at this point that the particularity of each theory enters the scene. We can have theories such that one can always find shifts such that the boundary term $A_\infty$ vanishes, and theories where this is not possible.%
\footnote{Of course, in the presence of a pole at $z\to \infty$ one can still write down recursion relations as long as 
one knows what the boundary term $A_{\infty}$ is. See references \cite{Feng1,Feng2,Benincasa} for work on this important issue.}
We defer the discussion of this important point for a moment, and for the time being assume that such a boundary term is absent. In this case, \eqref{sop} expresses $A(0)$, i.e.~the amplitude we wish to calculate, as a sum of residues. Using \eqref{h2} and the condition  $A_{\infty}=0$ we recast \eqref{sop} as 
\be
\label{ampli}
A (0) \ = \ \sum_{z_{ij}; h}  A_{L}^{h} (z= z_{ij}) {i\over P_{ij}^2}  A_{L}^{-h} (z= z_{ij}) \ .
\eeq
This is the BCFW recursion relation. It expresses the (unshifted) amplitude as a sum of products of two  amplitudes evaluated at shifted, complex kinematics. The fact that two amplitudes occur in the recursion is inherited from the factorisation of the amplitudes, which itself is a consequence of unitarity. 
The sum is over all possible arrangements of particles such that only one of the two shifted momenta belongs to either the left-hand or right-hand amplitude. With colour-ordered amplitudes, this corresponds to a simple sum, as illustrated in Figure \ref{rec_final_amplitude}.

Before concluding this section, we observe that the BCFW recursion relations had been originally found by a different route 
in \cite{bcfrec}. In that approach, one starts from a particularly simple and interesting combination of infrared consistency conditions found in 
\cite{Roiban:2004ix}. This infrared equation expresses a tree-level amplitude in $\cN=4$ super Yang-Mills as a linear combination of the coefficients of the one-mass and two-mass hard coefficients in the expansion of the corresponding one-loop amplitude. These coefficients can then be written using  generalised unitarity \cite{bcfgen}, and further elaborated using the explicit expressions of the three-point amplitudes appearing in their expression \cite{bcfrec}.

\subsection{Examples in Yang-Mills theory} 
Our first example of an application of the BCFW recursion relation will be the calculation of the colour-ordered gluon amplitude  $A(1^- 2^+ 3^- 4^+)$ in Yang-Mills theory. The first point to discuss is the choice of the shifts. We can understand this from the Feynman diagram point of view. When we perform shifts we will produce a certain sequence of, say $m$, propagators, between the shifted particles which depend on $z$. Their arguments  are of  the form $1/(P + z \eta)^2 $, which for large $z$ behaves as $[(P+ z \eta)^2]^{-m}  \sim z^{-m}$. 
Among all the diagrams, those which are leading at large $z$ are those where only three-point vertices appear, each of which providing a single positive power of momentum -- and hence of $z$ at large $z$. There will be exactly $m+1$ such vertices, giving a factor of $z^{m+1}$ at large $z$. Together with the propagators, this diagram behaves as $z$ at large $z$. In order to obtain a scattering amplitude we have to contract external lines with polarisation vectors. For gluons these are of the form 
\beq\label{gluon-pol}
\eps_{a \dot{a}}^{+} :=  
{\lt_{\dot{a}} 
\eta_a  \over \langle \eta \l \rangle  } \ , \qquad 
\eps_{a \dot{a}}^{-} := {\l_a\tilde{\eta}_{\dot a}  \over [\lt \tilde{\eta} ]} 
\ . 
\eeq
The only polarisation vectors which depend on $z$ are those of the shifted particles. It is clear that if we shift the 
holomorphic spinors $\l$ of a positive-helicity gluon, and the anti-holomorphic spinor 
$\lt$ of a negative-helicity gluon, these will provide a further suppression of $z^{-2}$ at large $z$. In conclusion, the diagram with the worst behaviour at large $z$ is suppressed as $z^{-1}$ at large $z$, on the condition that   we choose shifts as specified above.

Therefore, we will chose our shifts as
\beq
\label{YMshifts}
\hat{\lt}_1 := \lt_1 +  z \lt_2 \ , \qquad
\l_2 := \l_2 - z \l_1
\  ,
\eeq
with $\l_1$ and $\lt_2$ unshifted. In the literature such a shift is often denoted as a
$[12\rangle$ shift, and following the discussion from above we know that if legs 1 and 2 have helicities $-+$ the shifted amplitude $A(z)$ vanishes for large $z$ as required.
Using MHV diagrams, it can be shown \cite{bcfw} that the helicity choices $++$ and $--$ lead to the same benign large-$z$ behaviour.

\begin{figure}[ht]
\begin{center}
\scalebox{0.55}{\includegraphics{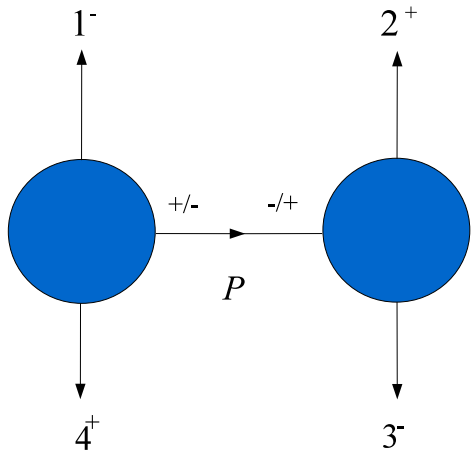}}
\end{center}
\caption{\it 
The two diagrams contributing to the Yang-Mills   amplitude 
$A  (1^- , 2^+ , 3^- , 4^+)$.  
The diagram with the internal helicity assignment $\langle +- \rangle$ vanishes because of the shifts \eqref{YMshifts}.}
\label{rec-YM}
\end{figure}

Now we move on to the calculation. 
There are two recursive diagrams to evaluate, depicted in Figure \ref{rec-YM}, differing in the helicity assignment of the internal gluon.  In both diagrams, $z$ is evaluated at the position of the pole $\lan \hat{2} 3\ran [23] = 0$, or 
\beq
\label{polez}
z =  {\lan 23\ran \over \lan 13 \ran} \ . 
\eeq
Consider first the diagram with internal helicities $\langle +- \rangle$. In this case the amplitude on the right-hand side has the MHV helicity configuration, and it is easy to see that it vanishes. Indeed, at the pole we have 
$\langle \hat{2} 3 \rangle = \langle  3  \hat{P} \rangle = \langle \hat{P} 2 \rangle = 0$. We now calculate the diagram with the opposite helicity assignment, $\langle -+ \rangle$. It is equal to 
\beq
\label{sssopra}
\left( 
 i  {\langle 1 \hat{P}  \rangle^3  \over  
 \langle \hat{P} 4 \rangle \langle 41 \rangle 
 } 
\right)\,
 {i\over P^2} 
 \left( 
 { -i
 [ - \hat{P} 2]^3 \over [23]  [3 -\!\!\hat{P}] } \right)
\ . 
\eeq
In our conventions, in which all particles have outgoing momenta, the spinors in the right-hand amplitude associated to the internal particle with on-shell momentum $-\hat{P}$ are%
$| - \hat{P}] = i |\hat{P}]$. 
We also have 
\beq
\label{up}
\langle1 \hat{P} \rangle [\hat{P} 2] = \langle1 |  \hat{P} | 2] =\langle1 |  p_2 + p_3 - z \l_1 \lt_2  | 2] = \lan13\ran [32] \ . 
\eeq
In principle in \eqref{up} we would have to insert the value \eqref{polez} of $z$ at the pole, but in this particular calculation the shift drops out since $[22]= 0$.  
Further multiplying and dividing by a factor of $\langle1 \hat{P} \rangle [\hat{P} 2] $ we form in the denominator the combinations 
\beqa
\langle1 \hat{P} \rangle [3 \hat{P} ] = - \langle1|p_2 + p_3 - z \l_1 \lt_2 | 3] = - \lan 12\ran [23] \ , 
\nonumber \\ 
 \langle \hat{P} 4 \rangle [\hat{P} 2  ]  = - \lan 4 | p_2 + p_3 - z \l_1 \lt_2 | 2] = - \lan 43 \ran [32] \ . 
 \eeqa
Collecting terms, \eqref{sssopra} becomes 
\beq
- i { \lan 13\ran^4 [32]^4  \over \lan 41\ran [23] \lan12 \ran [23] \lan43 \ran [32]} {1\over  \lan23\ran [32]}  \ = \ 
i {     \lan13\ran^4 \over \lan12\ran \lan23\ran\lan34\ran\lan41\ran }
\ , 
\eeq
which is the expected result. We remark that at no point in this derivation we needed to plug in the explicit value of $z$ in \eqref{polez}. This is due to the simplicity of this calculation; in general derivations one does need to insert the explicit value of $z$ at the poles for each diagram.

\subsection{Extensions to other theories}
The derivation presented above is very general, therefore one might expect to be able to apply on-shell  recursion relations
in different theories.  Specifically, at three stages we have made reference to a specific theory:
\begin{itemize} 
\item[{\bf 1.}] We have considered colour-ordered amplitudes in Yang-Mills theory. 
\item[{\bf 2.}] We have considered massless particles.
\item[{\bf 3.}] We have assumed that $A(z) \to 0$ as $z\to \infty$. 
\end{itemize} 
Let us analyse these three points in turn. 
Firstly, colour-ordering is just a technical point. We could have just as well applied the recursion to complete amplitudes, and nothing would have changed except that the sum over all recursive diagrams would have become  a double sum. We have also considered shifts appropriate for massless particles, but one could well extend these to massive particles as shown in
\cite{Badger:2005zh,Badger:2005jv}. For example, if particle 1 is massless and particle 2 is massive, one can consider the shifts 
\be
\hat{p}_1^{\dot{a} a} = \l_1^a ( \lt_1  + z (\l_1 p_2))^{\dot{a}}  \ , \quad 
\hat{p}_2^{\dot{a} a} =p_2^{\dot{a} a} - z \l_1^a (\l_1 p_2)^{\dot{a} }
\ , 
\ee
where we shift by the complex null momentum
\be
\eta^{a \dot{a}} = \l_1^a (\l_1 p_2)^{\dot{a}} \ ,
\ee
which has the properties $\eta \cdot \eta = p_1 \cdot \eta = p_2 \cdot \eta = 0$
as for the massless case, which are needed to preserve the mass-shell conditions of the unshifted momenta. There is another possibility namely we could have shifted by
\be
\eta'^{a \dot{a}} = (p_2 \tilde{\l}_1)^{a} \tilde{\l}_1^{\dot{a}} \ .
\ee
As in the massless case, we now consider the deformed tree-level amplitude $A(z)$ and use the fact that this function can have only simple poles in $z$ coming from internal propagators $i/P(z)^2$ going on-shell. This can only occur if particles 1 and 2 are on opposite sides of the propagator so that $P=p_j + \ldots + p_n+p_1$ and $P(z) = P + z \eta$. The propagator becomes
\be
\frac{i}{P(z)^2-m^2} = \frac{i}{P^2-m^2 + 2 z P \cdot \eta} \ .
\ee
Assuming that $A(z)$ for large $z$ and using  Cauchy's theorem leads in a similar fashion as in the massless case to the recursion relation
\be
A = \sum_{j=4}^{n} A_L(z_j) \frac{i}{P_j^2-m^2}  A_R(z_j)
\ , 
\ee
where $P_j = p_j + \ldots + p_n+p_1$, and $z_j = (m^2-P_j^2)/(2 P_j \cdot \eta)$.
Of course, we could have picked any other pair of external particles and even selected two massless particles with the usual BCFW shifts, or two massive particles --  although this last option is rather inconvenient for practical calculations.

As a simple illustration, we now consider the case of four-point amplitudes with two massive scalars. In order to do this, we need the 
expressions of the three-point amplitudes with  two massive scalars and one gluon as seed amplitudes. These were determined from Feynman rules in \cite{Badger:2005zh}, 
\be\label{mass3pt}
A_3(p_1^\pm, k^+,p_2^\mp) =\frac{\langle \eta_1 | p_1 | k ]}{\langle \eta_1 k \rangle} \ , \ \ \
A_3(p_1^\pm, k^-,p_2^\mp) =-\frac{\langle k | p_1 | \eta_2 ]}{[ \eta_2 k ]}
\ ,
\ee
with $p_{1,2}^2=m^2$ and $k^2=0$. 
Here, $\eta_{1,2}$ are two reference spinors needed for the definition of the polarisation vectors, and it can easily be seen that the
massive three-point amplitudes (\ref{mass3pt}) are independent of the choice of $\eta_{1,2}$.

Let us now focus on $A(p_1^+,1^+,2^+,p_2^-)$, the  amplitude of two positive helicity gluons and two massive scalars. We choose to shift the two gluon momenta $k_{1,2}$
with $\eta = \l_2^a \tilde{\l}_1^{\dot{a}}$, hence $\l_2$ and $\tilde{\l}_1$ are unshifted. There is only one recursive diagram to be considered involving two three-point amplitudes (\ref{mass3pt}) with a massive scalar propagating between them. This yields
\be
A_3(p_1^+, \hat{1}^+, -\hat{P}^-) \frac{i}{P^2-m^2} 
A_3(\hat{P}^+,\hat{2}^+, p_2^-) \ .
\ee
Choosing reference vectors $\eta_1 = \hat{k}_2$ and $\eta_2 = \hat{k}_1$ for the two three-point amplitudes one arrives at
\be
-i\frac{\langle 2 | p_1 | 1 ] \langle \hat{1}|p_2|\hat{2}]}{\langle12\rangle^2 ((p_1+k_1)^2-m^2)} \ ,
\ee
and using Fierz identities and on-shell conditions one can simplify this further to find
\be
A(p_1^+,1^+,2^+,p_2^-)=i\frac{m^2 [12]}{\langle 1 2 \rangle ((k_1+p_1)^2-m^2)}\ .
\ee
Similarly the amplitude with two massive scalars and two gluons of opposite helicity can be obtained from an almost identical recursion relation
\be
A(p_1^+,1^+,2^-,p_2^-)=-i\frac{\langle 2 |p_1|1]^2}{s_{12}((k_1+p_1)^2-m^2)}\ .
\ee
Other tree amplitudes involving massive scalars have been calculated explicitly in
\cite{Badger:2005zh,Badger:2005jv}. Such amplitudes have particular relevance for the calculation of one-loop amplitudes in QCD using unitarity. Due to the presence of rational terms, already mentioned at the end of Section \ref{rara}, it is not sufficient to consider unitarity cuts in four dimensions  -- these have to be performed in $D=4 - 2 \eps$ dimensions, if dimensional regularisation is used. If the particle running in the loop is a scalar, then going to $D$ dimensions effectively turns a massless scalar into a massive four-dimensional scalar  whose mass squared $\mu^2$ has to be integrated over ($\mu^2$ is identified with the $-2 \eps$ dimensional part of the loop momentum). Applications of this approach to calculations of one-loop QCD amplitudes are described in more detail in the chapter of this review from Britto \cite{britto}.


\subsection{Recursion relation in Einstein gravity}
It is very  interesting to consider on-shell  recursion relations in theories other than Yang-Mills, for instance Einstein gravity. Following from the BCFW work, this was first done in 
\cite{bbstrec,cs}.  A general formula for the 
MHV amplitudes for graviton scattering was
first given in \cite{Berends:1988zp}. 
Choosing  particles 1 and 2 to carry negative helicity, 
while the remainder carry positive helicity, the result of   \cite{Berends:1988zp} is   
\beqa
\label{BerendsGiele}
M (1, 2, 3, \cdots , n) =
i\langle 12\rangle^8 \Biggl[ {[12] [n-2 ~n-1] \over \langle 1~n-1\rangle}
{1\over N(n)} \prod_{i=1}^{n-3}
\prod_{j=i+2}^{n-1} \langle ij\rangle ~F
\ +\ {\cal P} (2,\ldots ,n-2)\Biggr],
\cr
\label{15}
\eeqa
where
\beq
F= \left\{ \begin{array}{ll}
\prod_{l=3}^{n-3}\,  [l| (p_{l+1} + p_{l+2}+\cdots
+p_{n-1}) | n\rangle &
n\geq 6\\
1& n=5\\
\end{array}
\right.\label{16}
\eeq
and ${\cal P} $ indicates permutations. The recursion relations for gravity on the other hand
yield the following solution for the MHV amplitudes
\cite{bbstrec}, 
\beqa
\label{npoint}
M(1, 2, {i_1}, \cdots , {i_{n-2}}) &=&
i\frac{\lan 1\, 2 \ran^6 [1\,  i_{n-2}]}{\lan 1 \, i_{n-2}\ran}\, \, 
G(i_1,i_2,i_3)\,
\prod_{s=3}^{n-3} \frac{\lan 2 | i_1+...+i_{s-1} | i_s]}
{\lan i_s i_{s+1}\ran \lan 2 i_{s+1}\ran }
\nonumber \\ 
&+& \cP(i_1,...,i_{n-2}),
\eeqa
where
\beq
G(i_1,i_2,i_3) = 
{1\over 2} 
\frac{[i_1 i_2]}{\lan 2i_1\ran\lan 2i_2\ran
\lan i_1i_2\ran\lan i_2i_3\ran\lan i_1i_3\ran}
\ .
\eeq
For $n=5$ the product term is dropped from \eqref{npoint}.

This is with the choice of reference 
legs $1^-$ and $2^-$. 
Subsequent work has expanded on these results, finding new
and interesting formulae  for the MHV amplitudes and proving equivalences between the different results
\cite{Nair:2005iv, Elvang:2007sg, Mason:2008jy, Nguyen:2009jk, Spradlin:2008bu, Feng:2010hd}. 
There are intriguing links to Yang-Mills theory -- see for example  \cite{Bern:2008qj} and references
therein -- and these are the subject of on-going research (see \cite{Feng:2010hd, Bern:2010fy, BjerrumBohr:2010yc} and references therein).

\begin{figure}[ht]
\begin{center}
\scalebox{0.55}{\includegraphics{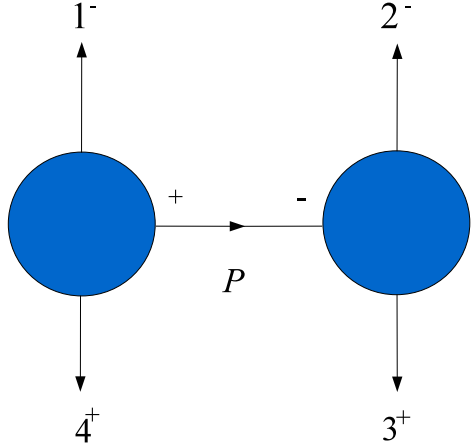}}
\end{center}
\caption{\it 
One of the two diagrams contributing to the recursion relation 
for the MHV  amplitude 
$M  (1^- , 2^- , 3^+ , 4^+)$.  
The other is obtained from this by swapping legs $3$ and  $4$.}
\label{recgrav}
\end{figure}

We would like to illustrate how recursion relations can be applied to gravity in a simple example. We will focus here on the four-point MHV amplitude of gravitons which  reads 
\beq
\label{4ptmhvgr}
M(1^-,2^-,3^+,4^+) \ = \ 
i\, \frac {\lan 1 2 \ran^8 [12] } {N(4) \lan 34 \ran}
\ ,
\eeq
where 
$ N(n)  :=   
\prod_{1\leq i < j \leq n} \lan i\, j\ran$.
We will consider the shifts 
\beq
\label{shiftsonceagain}
\hat{\l}_1 := \l_1 +  z \l_2 \ , \qquad
\hat{\lt}_2 := \lt_2 - z \lt_1
\  ,
\eeq
with $\l_2$ and $\lt_1$ unshifted. It is immediate to see that, under these shifts, one 
has
$M(\hat{1}^-,\hat{2}^-,3^+,4^+)\sim z^{-2}$ as $z \to \infty$,
hence there is no boundary term in the recursion relation. 
There are two   diagrams to consider, one of which is depicted 
in Figure \ref{recgrav}. The other is obtained by 
swapping the labels $4$ with $3$. 
The three-point amplitude on the left-hand side of Figure \ref{recgrav} has the $\overline{\rm MHV}$ helicity configuration, whereas the amplitude on the right-hand side has the MHV configuration. There is in principle another diagram where the left-hand amplitude is MHV and the right-hand one is $\overline{\rm MHV}$, but this  is easily seen to vanish, since the right-hand  $\overline{\rm MHV}$  factor would be  proportional to 
\beq
[3\, \hat{P} ] \, = \,
{ [ 3 | \hat{P} | 2 \ran \over \lan \hat{P} \, 2 \ran}
\, = \,
{ [ 3 | P | 2 \ran \over \lan \hat{P} \, 2 \ran} \, = \, 0
\ .
\eeq
We now evaluate explicitly the diagram in Figure \ref{recgrav}. It is equal to  
\beq
M^{(4)}_{\rm rec} (1,2,3,4) \ = \ M^{(3)} (\hat{P}^+ , 4^+, \hat{1}^-)   \, {i \over P^2} \, M^{(3)}  (-\hat{P}^- , \hat{2}^-, 3^+) \ , 
\eeq
where
\beqa
M^{(3)}   (\hat{P}^+ , 4^+, \hat{1}^-) & = & 
\left( -i
{ [ \hat{P} \, 4]^3 \over [ 4 \, 1] [ 1 \, \hat{P} ] }
\right)^2 \ , 
\\ \nonumber \cr
M^{(3)}  (P^- , \hat{2}^-, 3^+) & = & 
\left( i
{ \lan  - \hat{P} \, 2\ran^3 \over \lan 2 \, 3\ran  
\lan 3 \, -\hat{P} \ran } 
\right)^2 \ , 
\eeqa
and $P^2 = (p_1 + p_4)^2$.
Using 
$ \lan i \, \hat{P} \ran  =  
\lan i | P | 1] /  [ \hat{P} \, 1 ]$,  we get
\beq
M^{(4)}_{\rm rec} (1,2,3,4)  \ = \ 
i\, \frac {\lan 1 2 \ran^6 [14] } {\lan 14\ran \lan 23\ran^2
\lan 34\ran^2}
\ . 
\eeq
The full amplitude is $M^{(4)} (1^-,2^-,3^+,4^+) = M^{(4)}_{\rm rec} (1,2,3,4) + 
 M^{(4)}_{\rm rec} (1,2,4,3) $. 
We conclude that the result of the recursion relation  is
\beq
\label{sooo}
M(1^-,2^-,3^+,4^+)\  = \ 
i\, \frac {\lan 1 2 \ran^6 [14] } {\lan 14\ran \lan 23\ran^2
\lan 34\ran^2}
\, + \, 3 \leftrightarrow 4 
\ . 
\eeq
It is easy to check that \eqref{sooo} agrees 
with \eqref{4ptmhvgr}, thus reproducing correctly the expected four-point MHV amplitude. More examples of applications of the recursion relation are given in \cite{bbstrec,cs}. A thorough analysis of the large-$z$ behaviour of gravity amplitudes is performed in \cite{Benincasa:2007qj,ArkaniHamed:2008yf}.


\subsection{Manifestly supersymmetric recursion relations in $\cN=4$ SYM}
It is very natural to formulate manifestly supersymmetric on-shell recursion relations, and this is straightforward to write down  \cite{bhtrec,ahck}. The interest in such generalisations stems from the fact that having an algorithm that calculates amplitudes while respecting all symmetries of the theory is bound to produce expressions that are simple and compact. In particular, it was shown in \cite{bhtrec} using the supersymmetric recursion relations described below that the tree-level S-matrix of $\cN=4$ SYM is covariant under dual superconformal symmetry and that this covariance is manifest term-by-term in the recursion relation.

The main idea consists in supersymmetrising the momentum shifts of the standard BCFW recursion by accompanying them with corresponding supermomentum shifts. Consider again the shifts in \eqref{bcfshifts}. There we shifted $\lt_1$ and $\l_2$. The sum of the  supermomenta of the shifted particles, $q_1 + q_2 := \eta_1 \l_1 + \eta_2 \l_2$  should also be invariant under the shifts, hence the shift of $\l_2$, $\hat{\l}_2 = \l_2 - z \l_1$,  must be compensated by a corresponding shift of $\eta_1$, namely 
$\hat{\eta}_1 (z) = \eta_1+ z \eta_2$. Therefore, the set of (super)shifts
\beq
\label{supershift}
\hat{\lt}_1 (z) := \lt_1 + z \lt_2 \ , \quad
\hat{\l}_2 (z) = \l_2 - z \l_1 \ , \quad 
 \hat{\eta}_1 (z) = \eta_1+ z \eta_2 \ , 
 \eeq
preserves momentum and supermomentum conservations, as well as the on-shell conditions. 
This produces a one-parameter family of superamplitudes, 
\beq
\cA (z) := \cA (\{\l_1, \hat{\lt}_1, \hat{\eta}_1\}; \{\hat{\l}_1, \lt_2, \eta_1\}; \ldots )
\ , 
\eeq
where the dots denote the unshifted momenta and supermomenta for the remaining $n-2$ particles. 
The supersymmetric recursion relation follows  from arguments similar to those which 
led to \eqref{ampli}. We have 
\beq
\label{superampli}
\cA \ = \ 
\sum_{P} 
\int\!\!d^4\eta_{\hat{P}} \  \cA_L(z_{P})
 \frac{i}{P^2} \cA_R(z_{P})
 \ , 
\end{equation}
where $\eta_{\hat{P}}$ is the Grassmann  variable associated to the internal, 
on-shell leg with momentum $\hat{P}$.  The sum is over  all possible $P$ such that precisely 
one of the shifted momenta, say $\hat{p}_1$,  is contained in $P$. The two superamplitudes are then evaluated at the solution $z_P$ of the equation $\hat{P}^2 (z)=0$, where $\hat{P} (z) = P + z \l_1 \lt_2$. 

Notice that  superamplitudes are characterised by the number of external particles and 
their  total helicity, which is the sum of the helicities of all external particles. 
Hence, in the recursion relation \eqref{superampli}
we have an important constraint on  $\cA_L$ and  $\cA_R$, namely the
total helicity of $\cA_L$ plus the total helicity of $\cA_R$ must equal the total helicity of the
full amplitude $\cA$. This condition replaces the sum over internal helicities in the
standard  BCFW recursion  relation \eqref{ampli}.

In writing \eqref{superampli} we have assumed that $\cA(z) \to 0$ as $z\to \infty$, as was demonstrated in 
\cite{bhtrec, ahck}. Here we will present the approach of \cite{ahck}, showing that   
\beq
\label{behsup}
\cA_{\cN=4} (z) \sim {1\over z} \ , \quad 
\cA_{\cN=8} (z)\sim {1\over z^2} \ , \quad {\rm as} \ z\to \infty
\ .
\eeq
The key point is that one can use maximal supersymmetry 
to set to zero two of the $\eta^A$ variables, say 
$\eta_1$ and $\eta_2$,  in a given superamplitude $\cA( \l_1, \lt_1, \eta_1; \l_2, \lt_2, \eta_2; \cdots ; \l_n, \lt_n, \eta_n)$ in maximally supersymmetric $\cN=4$ super Yang-Mills or $\cN=8$ supergravity. 
Indeed, consider the particular combination of $\bar{q}$ supersymmetries defined by $\bar{q}_\zeta := \zeta^{\dot{\a}}_B \bar{q}^B_{\dot{\alpha}}$, where $B=1, \ldots , \cN$ and determine the $2 \cN$ parameters $ \zeta^{\dot{\a}}_B$ by requiring that 
\beq
e^{\bar{q}_\zeta} \eta_1^A = e^{\bar{q}_\zeta} \eta_2^A = 0
\ , 
\eeq
or, equivalently, 
$\bar{q}_\zeta \eta_{1,2}^A = - \eta_{1,2}^A $.
One easily finds that 
\beq
\zeta_{\dot\a}^A \ = {1\over [12]} \left(- \lt_{1; \dot\a} \eta_2^A +  \lt_{2; \dot\a} \eta_1^A\right)
\ , 
\eeq
hence the action on a generic coordinate $\eta_i$ is 
\beq
\label{etap}
e^{\bar{Q}_\zeta}  \eta_i \ := \ \eta_i^\prime \ = \ \eta_i - \eta_1 { [i2]\over [12]} + \eta_2 {[i1] \over [12]}
\ , 
 \eeq
and in particular $e^{\bar{Q}_\zeta}  \eta_1 = e^{\bar{Q}_\zeta}  \eta_2 \  = 0$. 

Invariance under $\bar{q}$ supersymmetry of an arbitrary  superamplitude $\cA$, which we write as  
\beq
\cA := \delta^{(4)} (p) \delta^{(2\cN)} (q) A(  \l_1, \lt_1, \eta_1; \l_2, \lt_2, \eta_2; \cdots ; \l_n, \lt_n, \eta_n)
\ , 
\eeq
implies that $\delta^{(4)} (p) \delta^{(2\cN)} (q) \, [\bar{q}^B_{\dot\beta} A] = 0 $, hence 
(omitting for brevity the $\delta^{(4)} (p) \delta^{(2\cN)} (q) $ factors) $e^{\bar{q}_\zeta} A = A$. Acting with the $\bar{q}$ operator explicitly, we therefore find that 
 \beqa
 \label{remove}
&& \cA (\l_1, \lt_1, 0; \l_2, \lt_2, 0; \cdots ;\l_i , \lt_i, \eta_i^\prime; \cdots ; \l_n, \lt_n, \eta_n^\prime) 
\nonumber \\ 
 &=&\cA (\l_1, \lt_1, \eta_1; \l_2, \lt_2, \eta_2; \cdots; \l_i , \lt_i, \eta_i; \cdots ; \l_n, \lt_n, \eta_n)
 \ , 
 \eeqa
where $\eta_i^\prime$ is given in \eqref{etap}. 

The application of the above to prove the large-$z$ behaviour of superamplitudes works as follows \cite{ahck}. Consider the family of amplitudes $\cA (z)$ obtained by performing the supershifts in \eqref{supershift}, and use \eqref{remove} to set $\eta_1 (z) $ and $\eta_2 (z)$ to zero. Importantly, the necessary supersymmetry parameter turns out to be $z$-independent: 
\beq
\zeta_{\dot\a}^A \ = \  {1\over [12]} \left(- \hat{\lt}_{1; \dot\a} \eta_2^A +  \lt_{2; \dot\a} \hat{\eta}_1^A\right) \ = \ 
  {1\over [12]} \left(- {\lt}_{1; \dot\a} \eta_2^A +  \lt_{2; \dot\a} {\eta}_1^A\right)
\ , 
 \eeq
where we used \eqref{supershift}. This implies that 
\beq
\label{abb}
\cA(z) \ = \ \cA( \l_1, \hat{\lt}_1, 0; \hat{\l}_2, \lt_2, 0; \cdots ; \l_i, \lt_i, \eta_i^\prime; \cdots ; \l_n, \lt_n, \eta_n^\prime) 
\ , 
\eeq
where none of the Grassmann variables $\eta_i^\prime$ in \eqref{abb} contain $z$, i.e.~the only $z$-dependence occurs through $\hat{\lt}_1$ and $\hat{\lt}_2$. As a consequence, the behaviour of  $\cA(z)$ at large $z$ will be the same of that of a gluon  amplitude in   Yang-Mills, or graviton amplitude had we started with $\cN=8$ supergravity,  
with gluons/gravitons 1 and 2 both with positive helicity (since we have set to zero the corresponding $\eta$ variables for particles 1 and 2). In the Yang-Mills case, such amplitudes are known to fall off as $1/z$ at large $z$ \cite{bcfw}, whereas in  gravity  the fall off is $1/z^2$ \cite{ArkaniHamed:2008yf},%
\footnote{Some of this benign large-$z$  behaviour, which is not apparent from a  Feynman diagram analysis, was observed already in \cite{bbstrec, cs,Benincasa:2007qj}. See also \cite{Cheung:2008dn} for a more recent investigation of the large-$z$ behaviour of scattering amplitudes in various theories. } 
 and \eqref{behsup} follows. 

 \subsection{Example: MHV superamplitude}
\begin{figure}[ht]
\begin{center}
\scalebox{0.55}{\includegraphics{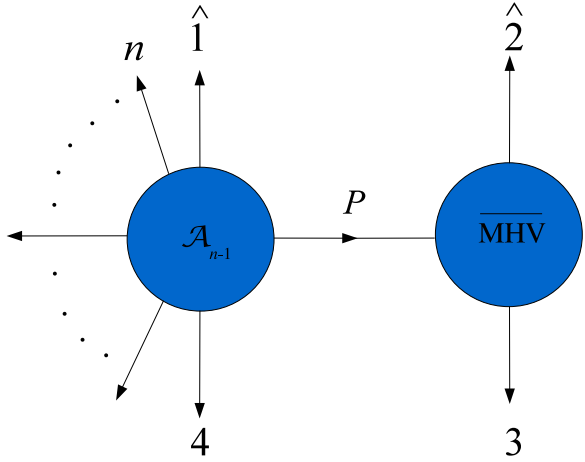}}
\end{center}
\caption{\it 
The single recursive diagram  contributing to the recursion relation 
for the $n$-point MHV  superamplitude. The amplitude on the right is anti-MHV, whereas that on the left is an $(n-1)$-point MHV superamplitude.}
\label{MHVsuper}
\end{figure}

 The simplest application of the supersymmetric recursion relation of   \cite{bhtrec,ahck} is to the derivation of the MHV superamplitude
 \beq
\label{NairMHV}
\cA_{\rm MHV} (1, \ldots , n) \ = \ i (2\pi)^4 \, 
{\delta^{(4)} (\sum_{i=1}^{n} \l_i \lt_i ) \, \delta^{(8)} ( \sum_{i=1}^{n} \eta_i \l_i ) 
\over \lan12\ran \cdots \lan n1\ran } 
\ .
\eeq
Performing the supershifts in \eqref{supershift}, we find that there is a single diagram contributing to the supersymmetric recursion relation, represented in Figure \ref{MHVsuper}.
 In this diagram, the right-hand amplitude is always a three-point ${\rm \overline{MHV}}$ superamplitude, whereas the amplitude on the left-hand side is {\rm MHV}.  The expression of the former was found in \cite{bhtrec, ahck} and is 
\beq
\label{3ptmhvbar}
\cA_{\overline{\mathrm{MHV}}} (1, 2, 3) \ = \ i (2 \pi)^4 \, \d^{(4)} (p_1 + p_2 + p_3 )\, 
{ \delta^{(4)} ( \eta_1 [23] \, + \, \eta_2 [ 31] \, + \, \eta_3 [12] ) \over [12] \, [23]\, [31] } 
\ . 
\eeq 
In \cite{bhtrec} it was also explicitly shown that this three point amplitude is invariant under supersymmetry, as well as covariant under the dual superconformal symmetry introduced in \cite{dhks}.

We describe in detail the derivation of the four-point superamplitude; the generalisation to higher numbers of points is entirely straightforward.   In this case the amplitude on the left-hand side is a three-point MHV superamplitude, 
\beq
\label{3ptmhv}
\cA_{{\mathrm{MHV}}} (1, 2, 3) \ = \ i (2 \pi)^4 \, \d^{(4)} (p_1 + p_2 + p_3 )\, 
{ \delta^{(8)} ( \eta_1 \l_1\, + \, \eta_2 \l_2 \, + \, \eta_3\l_3 ) \over \lan12\ran  \, \lan 23\ran \, \lan 31\ran } 
\ . 
\eeq 
Notice that    \eqref{3ptmhvbar}  and  \eqref{3ptmhv} can both be determined from symmetry considerations alone (up to an overall normalisation). For instance, 
\eqref{3ptmhv} is found  by requiring that it depends only on the holomorphic spinors $\l_1, \l_2, \l_3$ and satisfies the three equations 
\beq
\hat{h}_ i\,  \cA_{{\mathrm{MHV}}} (1, 2, 3) \ = \ \cA_{{\mathrm{MHV}}} (1, 2, 3) \ , \qquad i = 1, 2, 3 \ , 
\eeq
where the helicity operator for particle $i$ is \cite{witten}
\beq
\hat{h}_i \ :=  \ {1\over 2}  \left( 
-  \l_{i}^{\alpha} {\partial \over \partial \l_{i}^{ \alpha}}  +  
\lt_{i}^ { \dot{\alpha}}    {\partial \over \partial \lt_{i}^{\dot{\alpha}}} + 
\eta^A_i {\partial \over \partial \eta^A_i}\right)
\ . 
\eeq  
After this short detour, we now apply the supersymmetric recursion relation \eqref{superampli}, with 
 \beqa
\cA_L & = & \delta^{(4)} (p_4+  \hat{p}_1 + \hat{P} ) \, {
\delta^{(8)} (   q_4 +  \l_1 \hat{\eta}_1  +  \eta_{\hat{P}} \l_{\hat{P}}  ) 
\over 
\lan 1 \hat{P} \ran \lan \hat{P}  4 \ran \lan 4 1 \ran 
} 
\ , 
\\ \nonumber
\cA_R & = &  {\delta^{(4)} ( \hat{p}_2 + p_3 - \hat{P} ) \, \delta^{(4)} ( \eta_{\hat{P}} [23] + \eta_2 
[3  -\!\!\hat{P}] + \eta_3 [-\!\hat{P} \, 2]   ) \over
[-\!\hat{P}\, 2] [23] [3\, -\!\hat{P}]
} 
\ .
\eeqa
Using  the identity
\beqa 
&&\ \delta^{(8)} ( \hat{\eta}_1 \l_1 + \eta_4 \l_4 + \eta_{\hat{P}} \l_{\hat{P}}  ) \, 
\delta^{(4)} ( \eta_{\hat{P}} [23] + \eta_2 
[3  -\!\!\hat{P}] + \eta_3 [-\!\hat{P} \, 2]   )
\nonumber \\ 
&=& \ 
\delta^{(8)} \Big( \sum_{i\in L, R} \hat{\eta}_i \hat{\l}_i \Big) \, 
\delta^{(4)} ( \eta_{\hat{P}} [23] + \eta_2 
[3  -\!\!\hat{P}] + \eta_3 [-\!\hat{P} \, 2]   ) \ , 
\eeqa
as well as 
\beq
\sum_i \hat{\eta}_i \hat{\lambda}_i = \sum_i \eta_i \lambda_i \ , \qquad \sum_i \hat{p}_i = \sum_i p_i\ , 
\eeq 
we arrive at 
\beqa
\cA(1, 2, 3, 4) &=& 
i\, \delta^{(4)} \Big(\sum_{i\in L,R} p_i \Big) \,
\delta^{(8)} \Big( \sum_{i\in L, R} \eta_i \l_i \Big)  \, A(1,2,3,4)\ , 
\eeqa
where 
\beq
A \ = \ 
{1\over P_{23}^2} \, {1\over \lan 41\ran  [23] \, 
\lan 1 \hat{P}\ran \lan \hat{P} 4\ran [ \hat{P} 2 ] [ 3 \hat{P}] 
} \, 
\int\!d^4\eta_{\hat{P}} \, 
\delta^{(4)} ( \eta_{\hat{P}} [23] + \eta_2 
[3 \hat{P}] + \eta_3 [\hat{P} 2]   )\ . 
\eeq
It is easy to see that 
$\lan 1 \hat{P}\ran \lan \hat{P} 4\ran [ \hat{P} 2 ] [ 3 \hat{P}] =
\lan12\ran \lan34\ran [23]^2$, therefore we conclude that 
\beq
A (1,2,3,4) \ = \  { 1 \over \lan 1 2\ran \lan  23\ran\lan 34\ran\lan 41\ran}
\ . 
\eeq
We have thus  reproduced the expected supersymmetric MHV superamplitude \eqref{NairMHV} in the four-point case. 
The generalisation of the calculation showed above to  an $n$-point MHV superamplitude 
is very simple. The only difference is that 
the amplitude on the left-hand side of Figure \ref {MHVsuper}  will be an $(n-2)$-point MHV superamplitude. 
The algebra is  identical to that of the four-point example discussed above and leads to the
expected result \eqref{NairMHV}. 
 
We conclude with a few references to relevant other work.
In \cite{Drummond:2008cr}, an explicit solution to the $\cN=4$ supersymmetric recursion relation was found -- this is described elsewhere in this review. In $\cN=8$ supergravity it is also possible to present an explicit solution to the recursion relations, see \cite{Drummond:2009ge}. 
 Applications of recursion relations to rational terms in one-loop QCD have been considered in \cite{Bern:2005hs,Bern:2005ji,Bern:2005cq}, and are also discussed elsewhere in this volume. 
  The relation of recursion relations and twistor space has been  considered 
 in \cite{Hodges:2005bf,Hodges:2005aj,Hodges:2006tw,ArkaniHamed:2009si,Mason:2009sa}. Finally, recursion relations in three-dimensional supersymmetric Yang-Mills and Chern-Simons matter theories have recently been written down in \cite{Gang:2010gy, Agarwal}.

\vspace{0.7cm}
\section*{Acknowledgements}

It is a pleasure to thank James Bedford, Paul Heslop, Gang Yang  and Costas Zoubos for collaboration on the topics described in this article, and Radu Roiban and Gang Yang for a careful reading of a preliminary version of this article.  
This work was supported by the STFC under a Rolling Grant  ST/G000565/1.
WJS was supported by a Leverhulme Fellowship, and GT by an EPSRC Advanced Research Fellowship EP/C544242/1.


\end{document}